\documentclass[amsmath,amssymb,showpacs,aps,prd,superscriptaddress,twocolumn,floatfix,nofootinbib]{revtex4}

\usepackage{graphicx}
\usepackage{dcolumn}
\usepackage{bm}
\usepackage{subfigure} 
\usepackage{slashed} 

\newcommand{\be}{\begin{equation}} 
\newcommand{\ee}{\end{equation}}
\newcommand{\bea}{\begin{eqnarray}} 
\newcommand{\eea}{\end{eqnarray}}
\newcommand{\nin}{\noindent}
\newcommand{\nch}{N_{\text{ch}}}

\begin{document}



\title{Determination of the Atmospheric Neutrino Flux and \\
  Searches for New Physics with AMANDA-II}


\affiliation{III Physikalisches Institut, RWTH Aachen University, D-52056 Aachen, Germany}
\affiliation{Dept.~of Physics and Astronomy, University of Alabama, Tuscaloosa, AL 35487, USA}
\affiliation{Dept.~of Physics and Astronomy, University of Alaska Anchorage, 3211 Providence Dr., Anchorage, AK 99508, USA}
\affiliation{CTSPS, Clark-Atlanta University, Atlanta, GA 30314, USA}
\affiliation{School of Physics and Center for Relativistic Astrophysics, Georgia Institute of Technology, Atlanta, GA 30332, USA}
\affiliation{Dept.~of Physics, Southern University, Baton Rouge, LA 70813, USA}
\affiliation{Dept.~of Physics, University of California, Berkeley, CA 94720, USA}
\affiliation{Lawrence Berkeley National Laboratory, Berkeley, CA 94720, USA}
\affiliation{Institut f\"ur Physik, Humboldt-Universit\"at zu Berlin, D-12489 Berlin, Germany}
\affiliation{Universit\'e Libre de Bruxelles, Science Faculty CP230, B-1050 Brussels, Belgium}
\affiliation{Vrije Universiteit Brussel, Dienst ELEM, B-1050 Brussels, Belgium}
\affiliation{Dept.~of Physics, Chiba University, Chiba 263-8522, Japan}
\affiliation{Dept.~of Physics and Astronomy, University of Canterbury, Private Bag 4800, Christchurch, New Zealand}
\affiliation{Dept.~of Physics, University of Maryland, College Park, MD 20742, USA}
\affiliation{Dept.~of Physics and Center for Cosmology and Astro-Particle Physics, Ohio State University, 191 W.~Woodruff Ave., Columbus, OH 43210, USA}
\affiliation{Dept.~of Physics, TU Dortmund University, D-44221 Dortmund, Germany}
\affiliation{Dept.~of Subatomic and Radiation Physics, University of Gent, B-9000 Gent, Belgium}
\affiliation{Max-Planck-Institut f\"ur Kernphysik, D-69177 Heidelberg, Germany}
\affiliation{Dept.~of Physics and Astronomy, University of California, Irvine, CA 92697, USA}
\affiliation{Laboratory for High Energy Physics, \'Ecole Polytechnique F\'ed\'erale, CH-1015 Lausanne, Switzerland}
\affiliation{Dept.~of Physics and Astronomy, University of Kansas, Lawrence, KS 66045, USA}
\affiliation{Dept.~of Astronomy, University of Wisconsin, Madison, WI 53706, USA}
\affiliation{Dept.~of Physics, University of Wisconsin, Madison, WI 53706, USA}
\affiliation{Institute of Physics, University of Mainz, Staudinger Weg 7, D-55099 Mainz, Germany}
\affiliation{University of Mons-Hainaut, 7000 Mons, Belgium}
\affiliation{Bartol Research Institute and Department of Physics and Astronomy, University of Delaware, Newark, DE 19716, USA}
\affiliation{Dept.~of Physics, University of Oxford, 1 Keble Road, Oxford OX1 3NP, UK}
\affiliation{Dept.~of Physics, University of Wisconsin, River Falls, WI 54022, USA}
\affiliation{Dept.~of Physics, Stockholm University, SE-10691 Stockholm, Sweden}
\affiliation{Dept.~of Astronomy and Astrophysics, Pennsylvania State University, University Park, PA 16802, USA}
\affiliation{Dept.~of Physics, Pennsylvania State University, University Park, PA 16802, USA}
\affiliation{Dept.~of Physics and Astronomy, Uppsala University, Box 516, S-75120 Uppsala, Sweden}
\affiliation{Dept.~of Physics and Astronomy, Utrecht University/SRON, NL-3584 CC Utrecht, The Netherlands}
\affiliation{Dept.~of Physics, University of Wuppertal, D-42119 Wuppertal, Germany}
\affiliation{DESY, D-15735 Zeuthen, Germany}

\author{R.~Abbasi}
\affiliation{Dept.~of Physics, University of Wisconsin, Madison, WI 53706, USA}
\author{Y.~Abdou}
\affiliation{Dept.~of Subatomic and Radiation Physics, University of Gent, B-9000 Gent, Belgium}
\author{M.~Ackermann}
\affiliation{DESY, D-15735 Zeuthen, Germany}
\author{J.~Adams}
\affiliation{Dept.~of Physics and Astronomy, University of Canterbury, Private Bag 4800, Christchurch, New Zealand}
\author{M.~Ahlers}
\affiliation{Dept.~of Physics, University of Oxford, 1 Keble Road, Oxford OX1 3NP, UK}
\author{K.~Andeen}
\affiliation{Dept.~of Physics, University of Wisconsin, Madison, WI 53706, USA}
\author{J.~Auffenberg}
\affiliation{Dept.~of Physics, University of Wuppertal, D-42119 Wuppertal, Germany}
\author{X.~Bai}
\affiliation{Bartol Research Institute and Department of Physics and Astronomy, University of Delaware, Newark, DE 19716, USA}
\author{M.~Baker}
\affiliation{Dept.~of Physics, University of Wisconsin, Madison, WI 53706, USA}
\author{S.~W.~Barwick}
\affiliation{Dept.~of Physics and Astronomy, University of California, Irvine, CA 92697, USA}
% ...


\date{\today}

\collaboration{The IceCube Collaboration}
\homepage[]{http://www.icecube.wisc.edu}
\noaffiliation


\begin{abstract}
The AMANDA-II detector, operating since 2000 in the deep ice at the geographic South
Pole, has accumulated a large sample of atmospheric muon
neutrinos in the 100 GeV to 10 TeV energy range.  The zenith angle and
energy distribution of these events can be used to search for various
phenomenological signatures of quantum gravity in the neutrino sector, such
as violation of Lorentz invariance (VLI) or quantum decoherence (QD).  Analyzing a set
of 5511 candidate neutrino events collected during 1387 days of livetime from
2000 to 2006, we find no evidence for such effects and set upper limits on VLI
and QD parameters using a maximum likelihood method.  Given the absence of
evidence for new
flavor-changing physics, we use the same methodology to determine the
conventional atmospheric muon neutrino flux above 100 GeV.
\end{abstract}


\maketitle


\section{\label{sec:intro}Introduction}

Experimental searches for possible low-energy signatures of quantum gravity
(QG) can provide a valuable connection to a Planck-scale theory.  Numerous
quantum gravity theories suggest that Lorentz invariance may be violated or 
spontaneously broken, including loop quantum gravity \cite{gambini},
noncommutative geometry \cite{madore}, and string theory \cite{samuel}.
This, in turn, has encouraged phenomenological developments and 
experimental searches for such effects \cite{amelino05,mattingly05}.  Space-time may
also exhibit a ``foamy'' nature at the smallest length scales, inducing decoherence of
pure quantum states to mixed states during propagation through this
background \cite{hawking}.

The neutrino sector is a promising place to search for such phenomena.
Neutrino oscillations act as a quantum interferometer, and QG effects that
are expected to be small at energies below the Planck scale can be amplified into large
flavor-changing signatures.  Water-based or ice-based Cherenkov neutrino
detectors such as BAIKAL \cite{baikal}, AMANDA-II \cite{amanda_nature},
ANTARES \cite{antares}, and IceCube \cite{ahrens04} have the potential to accumulate large 
samples of high energy atmospheric muon neutrinos.
We present here an analysis of AMANDA-II atmospheric muon neutrinos
collected from 2000 to 2006 in which we search for flavor-changing
signatures that might arise from QG phenomena. 

In addition to searches for physics beyond the Standard Model, a measurement of
the conventional atmospheric neutrino flux is useful in its own right.
Uncertainties in the incident primary cosmic ray spectrum and in the
high energy hadronic interactions affect the atmospheric neutrino flux
calculations (see e.g.~Refs.~\cite{honda,barr_unc}).  Atmospheric neutrinos are the
primary background to searches for astrophysical neutrino point sources and diffuse
fluxes, so knowledge of the flux at higher energies is crucial.  In this
analysis, we vary the normalization and spectral index of existing models
for the atmospheric neutrino flux and determine the best-fit spectrum.  

We begin with a review of the phenomenology relevant to
our search for new physics in atmospheric neutrinos.  Next, we describe the
AMANDA-II detector, data selection procedures, and atmospheric neutrino
simulation.  Third, we describe the analysis methodology by which we
quantify any deviation from conventional physics.   We do not observe any
such deviation, and hence we present
upper limits on violation of Lorentz invariance (VLI) and quantum decoherence
(QD) obtained with this methodology, as well as a determination of the
conventional atmospheric neutrino flux.


\section{\label{sec:pheno}Phenomenology}
\subsection{Atmospheric Neutrinos}

Atmospheric neutrinos are produced when high energy cosmic rays collide
with air molecules, producing charged pions and kaons that subsequently
decay into muons and muon neutrinos.  Observations of atmospheric neutrinos
by Super-Kamiokande \cite{superk04}, Soudan 2 \cite{soudan03}, MACRO
\cite{macro02}, and other experiments have provided strong evidence for
mass-induced atmospheric neutrino oscillations.  The relationship between the mass
eigenstates and the flavor eigenstates can be characterized by
three mixing angles, two mass splittings, and a complex phase.  Because of
the smallness of the $\theta_{13}$ mixing angle and the $\Delta m_{12}$
splitting (see Ref.~\cite{global_osc_08} for a review), it suffices to consider a
two-neutrino system in the atmospheric case, and the survival probability for muon
neutrinos of energy $E$ as they travel over a baseline $L$ from the
production point in the atmosphere to a detector is

\begin{equation}
\label{osc_psurv}
  P_{\nu_{\mu} \rightarrow \nu_{\mu}} = 1\ -\ \sin^2 2\theta_{\text{atm}}\ \sin^2 \left(
  \frac{\Delta m^2_{\text{atm}}L}{4E}\right)\ ,
\end{equation}

\noindent where $L$ is in inverse energy units (we continue this
convention unless noted otherwise).  In practice, the zenith
angle of the neutrino serves as a proxy for the baseline $L$. 

A recent global fit to oscillation data results in
best-fit atmospheric oscillation parameters of $\Delta m^2_{\text{atm}} = 2.39 \times
10^{-3}\ \text{eV}^2$ and $\sin^2 2\theta_{\text{atm}} = 0.995$ \cite{global_osc_08}.  
Thus, for energies above about 50 GeV, atmospheric neutrino oscillations
cease for Earth-diameter baselines.  However, a number of
phenomenological models of physics beyond the Standard Model predict
flavor-changing effects at higher energies that can alter the zenith angle
distribution and energy spectrum of atmospheric muon neutrinos.  We review
two of these here, violation of Lorentz invariance and quantum decoherence.  

\subsection{Violation of Lorentz Invariance}

Many models of quantum gravity suggest that Lorentz symmetry may not be
exact \cite{mattingly05}.  Even if a QG theory is Lorentz symmetric,
the symmetry may still be spontaneously broken in our Universe.
Atmospheric neutrinos, with energies above 100 GeV and mass less than 1 eV,
have Lorentz boosts exceeding $10^{11}$ and provide a sensitive test
of Lorentz symmetry.  

Neutrino oscillations in particular provide a sensitive testbed for such
effects.  Oscillations act as a ``quantum interferometer'' by magnifying
small differences in energy into large flavor changes as the neutrinos
propagate. In conventional oscillations, this energy shift 
results from the small differences in mass among the eigenstates, but
specific manifestations of VLI can also result in energy shifts that can
generate neutrino oscillations with different energy dependencies.

In particular, we consider VLI in which neutrinos have limiting velocities
other than the canonical speed of light $c$ (\cite{coleman99,glashow04}; see the appendix for further
background).  Since these velocity eigenstates can be distinct from the mass or flavor
eigenstates, in a two-flavor system this introduces another mixing angle
$\xi$ and a phase $\eta$.  The magnitude of the VLI is characterized by the
velocity-splitting between the eigenstates, $\Delta c/c = (c_{a1} - c_{a2})
/c$. 

In this form of VLI, the $\nu_\mu$ survival probability is \cite{ggm04}

\begin{equation}
  \label{vli_psurv}
  P_{\nu_{\mu} \rightarrow \nu_{\mu}} = 1\ -\ \sin^2 2\Theta\ \sin^2 \left(
  \frac{\Delta m^2L}{4E}\ \mathcal{R}\right)\ ,
\end{equation}

\noindent where the combined effective mixing angle $\Theta$ can be written 

\begin{equation}
  \sin^2 2\Theta = \frac{1}{\mathcal{R}^2}(\sin^2 2\theta + R^2 \sin^2 2\xi
  + 2R\sin 2\theta \sin 2\xi \cos \eta)\ ,
\end{equation}

\nin the correction to the oscillation wavelength $\mathcal{R}$ is

\begin{equation}
  \mathcal{R} = \sqrt{1 + R^2 + 2R(\cos 2\theta \cos 2\xi + \sin 2\theta
    \sin 2\xi \cos \eta)}\ ,
\end{equation}

\nin and the ratio $R$ between the VLI oscillation wavelength and
mass-induced wavelength is 

\begin{equation}
  R = \frac{\Delta c}{c}\frac{E}{2}\frac{4E}{\Delta m^2}
\end{equation}

\noindent for a muon neutrino of energy $E$ and traveling over baseline
$L$.  For atmospheric neutrinos, we fix the conventional mixing angle $\theta =
\theta_{\text{atm}}$ and mass difference $\Delta m^2 = \Delta
m_{\text{atm}}^2$ to the global fit values determined in
Ref.~\cite{global_osc} of $\Delta m^2_{\text{atm}} = 2.2 \times
10^{-3}\ \text{eV}^2$ and $\sin^2 2\theta_{\text{atm}} = 1$.  For simplicity, the phase $\eta$ is often set to 0
or $\pi/2$.  For illustration, if we take both conventional and VLI 
mixing to be maximal ($\xi = \theta = \pi/4$), this
reduces to 

\begin{equation}
\label{max_prob}
  P_{\nu_{\mu} \rightarrow \nu_{\mu}} \mathrm{(maximal)} = 1\ -\ \sin^2 \left(
  \frac{\Delta m^2L}{4E} + \frac{\Delta c}{c}\frac{LE}{2}\right)\ .
\end{equation}

\begin{figure}
\resizebox{0.45\textwidth}{!}{\includegraphics{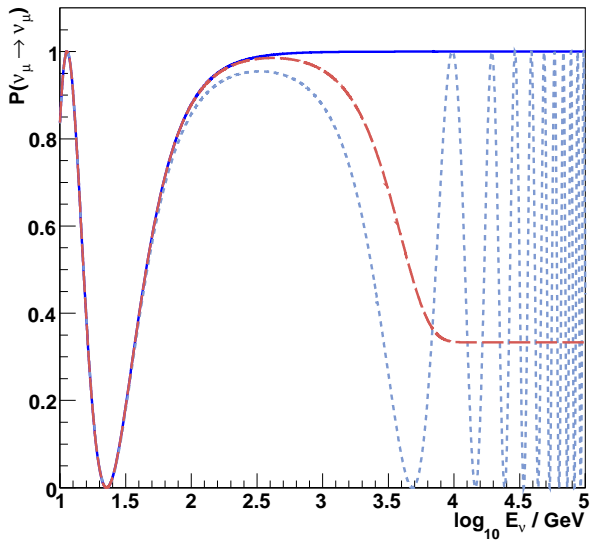}}
\caption{\label{fig:psurv} $\nu_\mu$ survival probability as a function of
  neutrino energy for maximal baselines ($L \approx
  2R_\text{Earth}$) given conventional oscillations (solid line), VLI
  (dotted line, with $n=1$, $\sin 2\xi = 1$, and $\Delta\delta = 10^{-26}$), and QD effects
  (dashed line, with $n=2$ and $D^* = 10^{-30}\ \text{GeV}^{-1}$).}  
\end{figure}

\nin Note the different energy dependence of the two effects.  The survival
probability for maximal baselines as a function of neutrino energy is shown in
Fig.~\ref{fig:psurv}.

Several neutrino experiments have set upper limits on this manifestation of
VLI, including MACRO 
\cite{macro_vli}, Super-Kamiokande \cite{superk_vli}, and a combined analysis
of K2K \cite{k2k} and Super-Kamiokande data \cite{ggm04} ($\Delta c/c < 2.0 \times
10^{-27}$ at the 90\% CL for maximal mixing).  In previous work, AMANDA-II has set
a preliminary upper limit using four years of data of $5.3 \times 10^{-27}$
\cite{amanda_vli_icrc}.  Other neutrino telescopes, such as ANTARES, are
also expected to be sensitive to such effects (see e.g.~Ref.~\cite{antares_vli}).   

Given the specificity of this particular model of VLI, we wish to
generalize the oscillation probability in Eq.\ \ref{vli_psurv}.  We follow
the approach in \cite{antares_vli}, which is to generalize the VLI
oscillation length $L \propto E^{-1}$ to other integral powers of the
neutrino energy $E$, that is, 

\be
\frac{\Delta c}{c}\frac{LE}{2} \rightarrow \Delta\delta \frac{LE^n}{2}\ ,
\ee

\nin where $n \in \{1,2,3\}$, and the generalized VLI term $\Delta\delta$ is in
units of $\mathrm{GeV}^{-n+1}$. An $L \propto E^{-2}$ energy dependence
($n=2$) has been proposed in the context of loop quantum 
gravity \cite{alfaro_le2} and in the case of non-renormalizable VLI effects
caused by the space-time foam \cite{brustein_le2}.  Both the $L \propto
E^{-1}$ ($n=1$) and the $L \propto E^{-3}$~$(n=3)$ cases have been examined
in the context of violations of the equivalence principle (VEP) \cite{gasperini_vep,
  halprin_vep, adunas_vep}.  In general, Lorentz violation
implies violation of the equivalence principle, so searches for either 
effect are related \cite{mattingly05}.  

\subsection{Quantum Decoherence}

Another possible low-energy signature of QG is the evolution of pure
states to mixed states via interaction with the environment of space-time
itself, or quantum decoherence.  One heuristic picture of this phenomenon
is the production of virtual black hole pairs in a ``foamy'' spacetime,
created from the vacuum at scales near the Planck length \cite{ellis84}.
Interactions with the virtual black holes may not preserve certain quantum
numbers like neutrino flavor, causing decoherence into a superposition of
flavors.

Quantum decoherence can be treated phenomenologically as a quantum open
system that evolves thermodynamically (we refer the reader to the appendix
for more detail).  In a three-flavor neutrino system, the decoherence from
one flavor state to a superposition of flavors can
be characterized by a set of parameters $D_i, i \in \{1,\ldots ,8\}$ that represent
a characteristic inverse length scale over which the decoherence sets in.
The $\nu_\mu$ survival 
probability in such a system is \cite{gago}

\begin{widetext}
\bea
\label{qd_psurv}
P_{\nu_{\mu} \rightarrow \nu_{\mu}} = \frac{1}{3} &+& \frac{1}{2} \Bigg[
\frac{1}{4} e^{-L D_3} (1 + \cos 2\theta)^2 +
\frac{1}{12} e^{-L D_8} (1 - 3\cos 2\theta)^2 + e^{-\frac{L}{2} (D_6+D_7)} \sin^2 2\theta \\
\nonumber & \cdot & 
\left(
\cos \left[\frac{L}{2} \sqrt{\left(\frac{\Delta m^2}{E}\right)^2 -
  (D_6-D_7)^2}\right]
+ \frac{\sin \left[\frac{L}{2} \sqrt{\left(\frac{\Delta m^2}{E}\right)^2 -
  (D_6-D_7)^2}\right](D_6-D_7)}{\sqrt{\left(\frac{\Delta m^2}{E}\right)^2 -
  (D_6-D_7)^2}}
\right)
\Bigg]\ .
\eea
\end{widetext}

\nin Note the limiting probability of $1/3$, representing full
decoherence into an equal superposition of flavors.  The $D_i$ not
appearing in Eq.~\ref{qd_psurv} affect decoherence between other flavors,
but not the $\nu_\mu$ survival probability.

The energy dependence of the decoherence terms $D_i$ depends on the
underlying microscopic model.  As with the VLI effects, we choose a generalized
phenomenological approach where we suppose the $D_i$ vary as some integral
power of the energy, that is

\be
\label{qd_energy}
D_i = D^*_i E^{n},\ n \in \{1,2,3\} 
\ee

\nin where $E$ is the neutrino energy in GeV, and the units of the
$D_i^*$ are $\mathrm{GeV}^{-n+1}$.  The particularly interesting
$E^2$ form is suggested by decoherence calculations in non-critical
string theories involving recoiling D-brane geometries \cite{dbrane}.  We
show the $n=2$ survival probability as a function of neutrino energy for
maximal baselines in Fig.~\ref{fig:psurv}.

An analysis of Super-Kamiokande in a two-flavor framework has
resulted in an upper limit at the 90\% CL 
of $D^* < 9.0\times10^{-28}\ \rm{GeV}^{-1}$ for an $E^2$ model and
all $D^*_i$ equal \cite{superk_qd}.  ANTARES has reported
sensitivity to various two-flavor decoherence scenarios as well, using a
more general formulation \cite{morgan_qd}.  Analyses of
Super-Kamiokande, KamLAND, and K2K data \cite{superk_k2k_qd, kamland_qd}
have also set strong limits on decoherence effects proportional to $E^0$
and $E^{-1}$.  Because for such effects our higher energy range does not
benefit us, we do not expect to be able to improve upon these limits, 
and we focus on effects with $n \ge 1$.


\section{\label{sec:amanda}Data and Simulation}

\subsection{The AMANDA-II Detector}

The AMANDA-II detector consists of 677 optical modules (OMs) on 19 vertical
cables or ``strings'' frozen into the deep, clear ice near the geographic
South Pole.  Each OM consists of a 20 cm diameter photomultiplier tube (PMT)
housed in a glass pressure sphere.  Cherenkov photons produced by
charged particles moving through the ice trigger the PMTs.  Combining the
photon arrival times with knowledge of the absorption and scattering
properties of the ice \cite{icepaper} allows reconstruction of a particle
track through the array \cite{amandareco}.  

In particular, a charged current $\nu_{\mu}$ interaction will produce a
muon that can traverse the entire detector.  This 
track-like topology allows reconstruction of the original neutrino
direction to within a few degrees. An estimate of the energy of the muon is
possible by measuring its energy loss, but this is complicated by
stochastic losses, and in any case is only a lower bound on the original
neutrino energy.

\subsection{Simulation}

In order to meaningfully compare our data with expectations from various
signal hypotheses, we must have a detailed simulation of the atmospheric
neutrinos and the subsequent detector response.  For the input atmospheric
muon neutrino spectrum, we generate an isotropic power-law flux with the 
 \textsc{nusim} neutrino simulator \cite{nusim} and then reweight the
 events to standard flux predictions \cite{bartol, honda}.  We have
 extended the predicted fluxes to the TeV energy range by fitting the
 low-energy region with the Gaisser parametrization \cite{gaisserbook} and
 then extrapolating above 700 GeV.  We add standard oscillations and/or
 non-standard flavor changes by weighting the events with the muon neutrino
 survival probability in Eqs. \ref{osc_psurv}, \ref{vli_psurv}, or
 \ref{qd_psurv}.   

Muon propagation and energy loss near and within the detector is simulated
using \textsc{mmc} \cite{mmc}.  Photon propagation through the ice, including
scattering and absorption, is modeled with \textsc{photonics}
\cite{photonics}, incorporating the depth-dependent characteristic dust
layers \cite{icepaper}.  The \textsc{amasim} program
\cite{amasim} simulates the detector response, and 
identical reconstruction methods are performed on data and simulation.
Cosmic ray background rejection is verified at all but the 
highest quality levels by a parallel simulation chain fed with atmospheric
muons from \textsc{corsika} \cite{corsika}, although when reaching
contamination levels of $O(1\%)$ --- a rejection factor of $10^8$ ---
computational limitations become prohibitive.   
 
\subsection{Atmospheric Neutrino Event Selection}

Even with kilometers of ice as an overburden, atmospheric
muon events dominate over neutrino events by a factor of about $10^6$.
Selecting only ``up-going'' muons allows us to reject the
large background of atmospheric muons, 
using the Earth as a filter to screen out everything but
neutrinos.  In practice, we must also use other observables indicating the
quality of the muon directional reconstruction, in order to eliminate
mis-reconstructed atmospheric muon events.  

Our data sample consists of $1.3\times 10^{10}$ events collected with
AMANDA-II during the years 2000 to 2006.  The primary trigger for this analysis is a
multiplicity condition requiring 24 OMs to exceed their discriminator
threshold (a ``hit'') within a sliding window of 2.5 $\mu$s.  As part of
the initial data cleaning, periods of unstable detector 
operation are discarded, such as during the austral summer months when
upgrades and configuration changes occur.  After accounting for inherent
detector deadtime in the trigger and readout electronics, the sample
represents 1387 days of livetime.  During  
the data filtering, dead or unstable OMs are removed, resulting in
approximately 540 modules for use in this analysis.  Isolated noise hits
and hits caused by electrical cross-talk are also removed \cite{amandareco}.

As a starting point for neutrino selection, we utilize the quality
selection criteria from the 
AMANDA-II 5-year point source analysis \cite{amanda5yr}.  These cuts,
not specifically optimized for high energy neutrinos, are efficient at
selection of atmospheric neutrinos and achieve a purity level of
$\sim95\%$, estimated by tightening the quality cuts until the ratio between data and
atmospheric neutrino simulation stabilizes.  The primary reconstruction
and/or quality variables used in this selection are:

\begin{enumerate}
\item the reconstructed zenith angle as obtained from a 32-iteration \emph{unbiased
  likelihood} (UL) fit;
\item the smoothness, a topological parameter describing the homogeneity of
  the photon hits along the UL fit track;
\item the estimated angular resolution of the UL fit, using the width of
  the likelihood minimum \cite{till};
\item the likelihood ratio between the UL fit and a \emph{Bayesian
  likelihood} (BL) fit \cite{bayes}, obtained by weighting the likelihood with a
  zenith-angle-dependent prior.  This weight constrains the track
  hypothesis to reconstruct the event as a ``down-going'' atmospheric muon.
\end{enumerate}

\nin The strength of the smoothness and the likelihood ratio cuts also vary
with the reconstructed zenith angle, as in general the cuts must be
stronger near the horizon where background contamination is worse.  Further
discussion of the background rejection of these quality variables can be
found in the point source analysis using these data \cite{amanda7yr}.

To this selection we add further criteria to remove the final few percent
of mis-reconstructed atmospheric muons.  Specifically, we remove events
with poor values in the following quality variables:

\begin{enumerate}
\item the space-angle difference between the UL fit track and the fit track
  by JAMS (a fast pattern-matching reconstruction; see Ref.~\cite{amanda5yr});
\item the number of hits from direct (unscattered) photons based on the UL fit
  hypothesis;
\item the maximum length along the reconstructed track between direct photon hits.
\end{enumerate}

\nin These selection criteria, as well as the analysis procedure described in section
\ref{sec:analysis}, were designed in a blind manner, in order to avoid
biasing the results.  Specifically, our observables (the zenith
angle and number of OMs hit, $\nch$; see section \ref{sec:observables})
were kept hidden when designing both.  However, after unblinding, we found a small excess
of high energy events above atmospheric neutrino predictions (444
events with $60 \le \nch < 120$ on an expectation of $\sim$350).
While this is a relatively small fraction of the overall sample, and
an excess at high $\nch$ cannot be misinterpreted as one of our new physics
hypotheses, a concentration of high energy background events could falsely
suggest an atmospheric neutrino spectrum much harder than expected.  

We find that these events exhibit characteristics of mis-reconstructed
atmospheric muons: poor reconstructed angular resolution; poor UL-to-BL likelihood
ratio; and low numbers of unscattered photon hits based on the fit
hypothesis.  As atmospheric neutrino events show better angular resolution
and likelihood ratio at higher energies, we chose to revise our selection
criteria to tighten the cuts on space-angle difference and angular resolution as
function of the number of OMs hit, $\nch$.  In particular, from $\nch=50$
to $\nch=80$, we linearly decrease (strengthen) the required angular resolution and
space-angle difference.  These additional cuts were 
only applied to events with likelihood ratio lower than the median for a
given zenith angle, as determined by atmospheric neutrino simulation.   We
estimate that the purity of the final event sample is greater than 99\%.

\subsection{Final Neutrino Sample}

After all selection criteria are applied, we are left with a sample of 5544 atmospheric
neutrino candidate events with reconstructed zenith angles below the
horizon\footnote{A table of the atmospheric neutrino events is available at
  \texttt{http://www.icecube.wisc.edu/science/data}\ .}.  We may characterize the
total efficiency of neutrino  
detection, including all detector and cut efficiencies as well as
effects such as earth absorption, via the neutrino
effective area $A^{\nu}_{\text{eff}}(E_{\nu},\theta,\phi)$, defined such
that

\be
\label{eq_aeff}
N_{\text{events}} = \int dE_{\nu}\ d\Omega\ dt\   
\frac{d\Phi(E_{\nu},\theta,\phi)}{dE_\nu d\Omega}\ A^{\nu}_{\text{eff}}(E_{\nu},\theta,\phi)
\ee

\begin{figure*}
\begin{center}
\subfigure{
\includegraphics[scale=0.41]{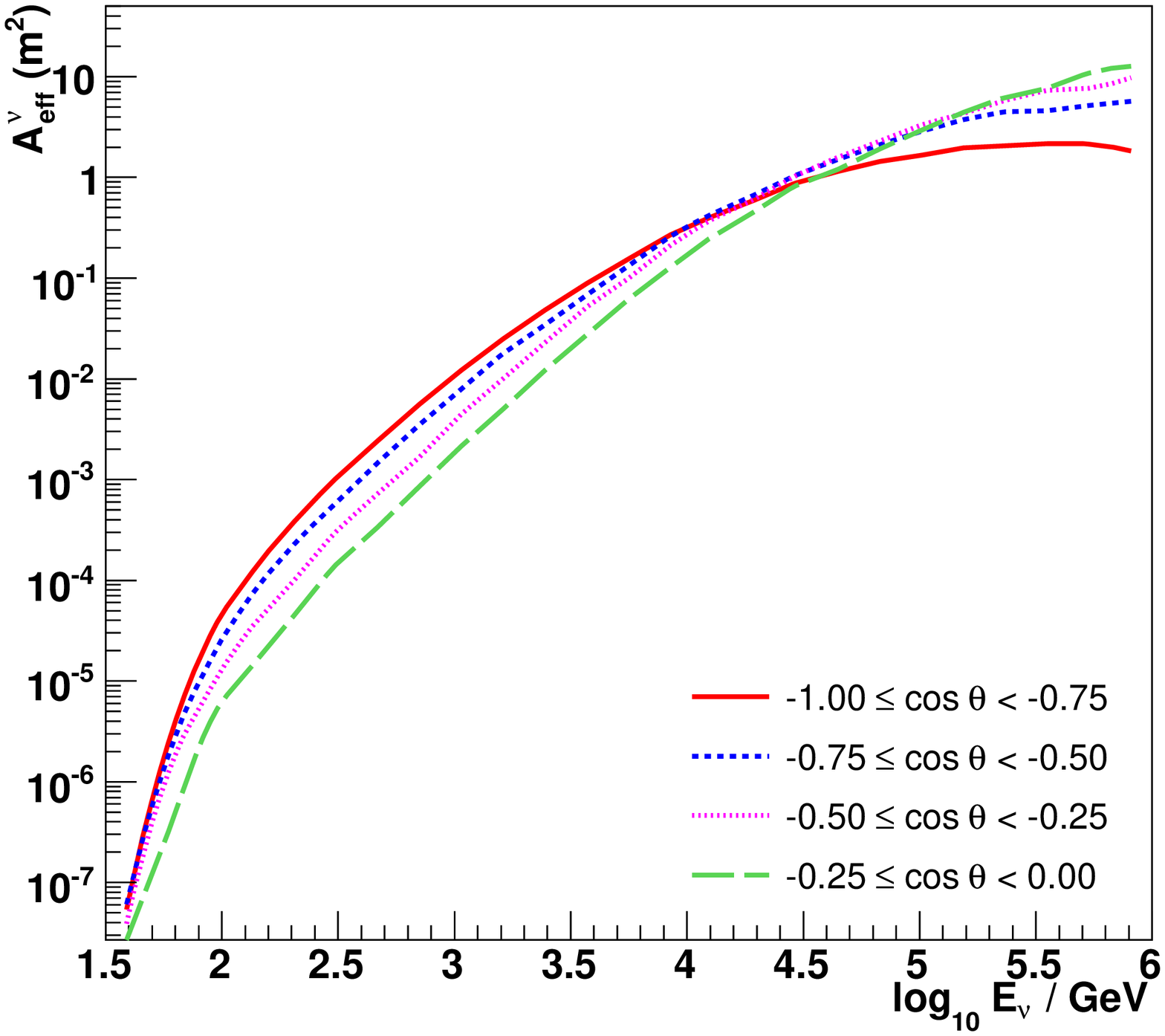}}%
\hfill
\subfigure{
\includegraphics[scale=0.39]{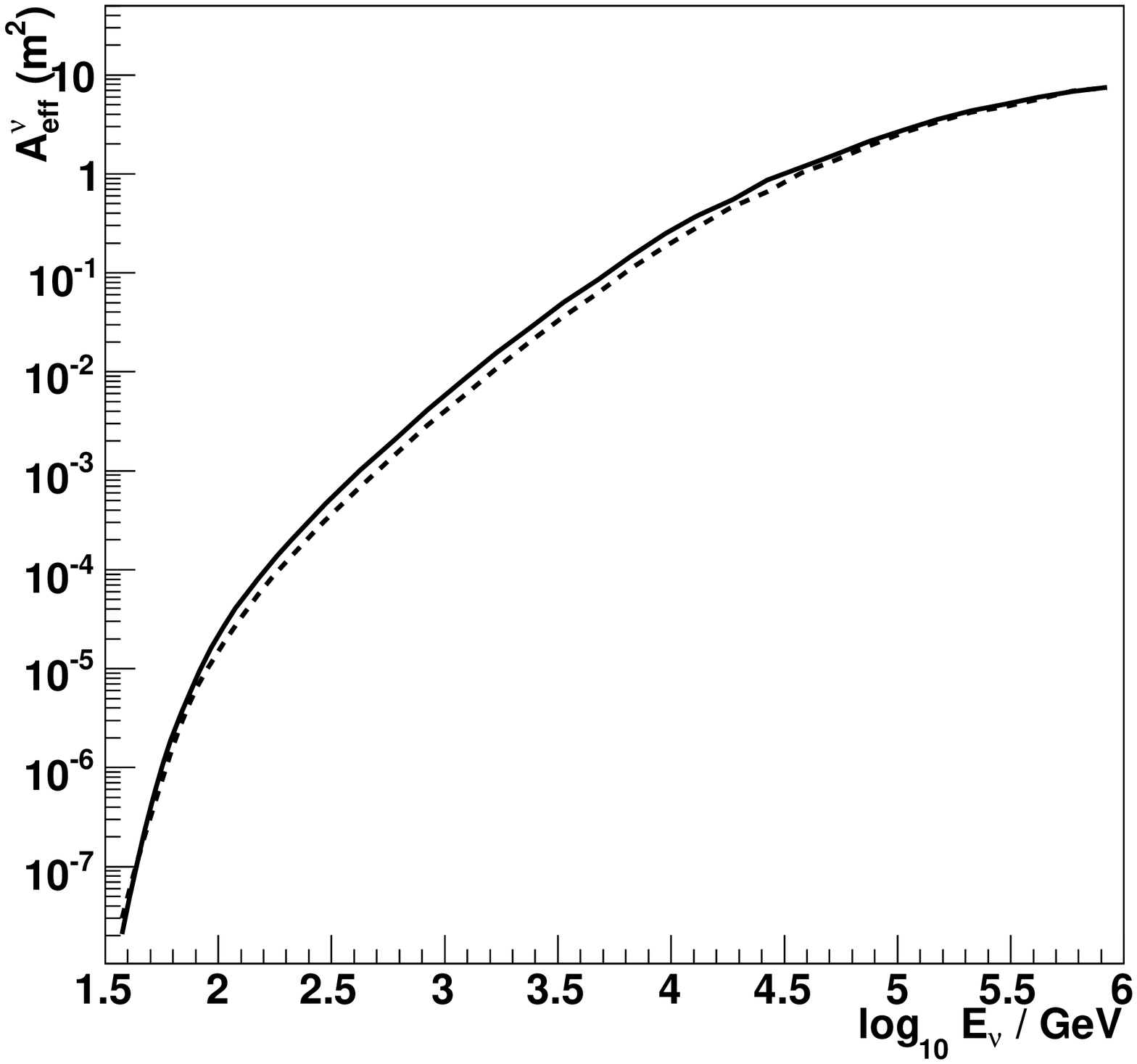}}%
\end{center}
\caption{\label{fig:aeff} Simulated detector effective area versus neutrino
  energy at the final analysis level.  Left: $\nu_\mu$ effective areas for several zenith angle
  ranges.  Right: zenith-angle-averaged effective areas for $\nu_\mu$
  (solid) and $\bar{\nu}_\mu$ (dotted).}
\end{figure*}

\nin for a differential neutrino flux $d\Phi/ dE_\nu d\Omega$.  
Fig.~\ref{fig:aeff} shows the $\nu_\mu$ and $\bar{\nu}_\mu$ effective areas
as a function of neutrino energy for event sample used in this analysis, as 
derived from the simulation chain described in the previous section.  We
have averaged over the detector azimuth $\phi$.  The differences in effective
area at various zenith angles are due to detector geometry, Earth
absorption at high energies, and the strong quality cuts near the horizon;
the different effective areas for $\nu_\mu$ and $\bar{\nu}_\mu$ are due to
their different interaction cross sections.

\begin{figure}
\resizebox{0.45\textwidth}{!}{\includegraphics{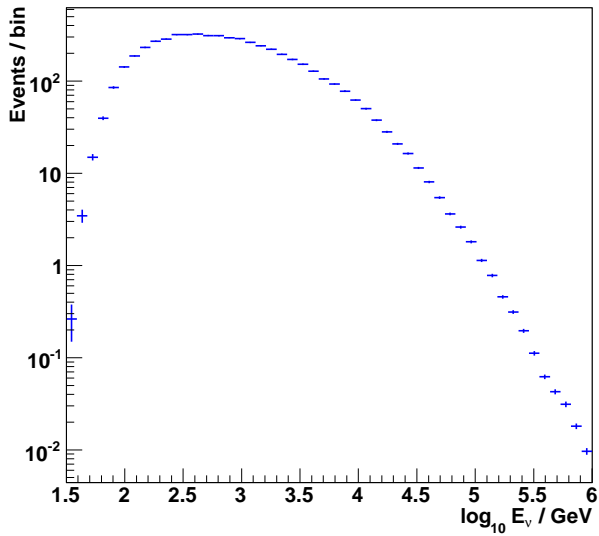}}
\caption{\label{fig:edist} Simulated $\nu_\mu+\bar\nu_\mu$ energy
  distribution of the final event sample, assuming the Barr \textit{et
    al.} input spectrum.} 
\end{figure}

The simulated energy response to the Barr
\textit{et al.}~atmospheric neutrino 
flux \cite{bartol} (without any new physics) is shown in Fig.~\ref{fig:edist}.  For
this flux, the simulated median energy of the final event sample is 640
GeV, and the 5\%-95\% range is 105 GeV to 8.9 TeV.

\section{\label{sec:analysis}Analysis Methodology}

\subsection{\label{sec:observables}Observables}

\begin{figure}
\resizebox{0.45\textwidth}{!}{\includegraphics{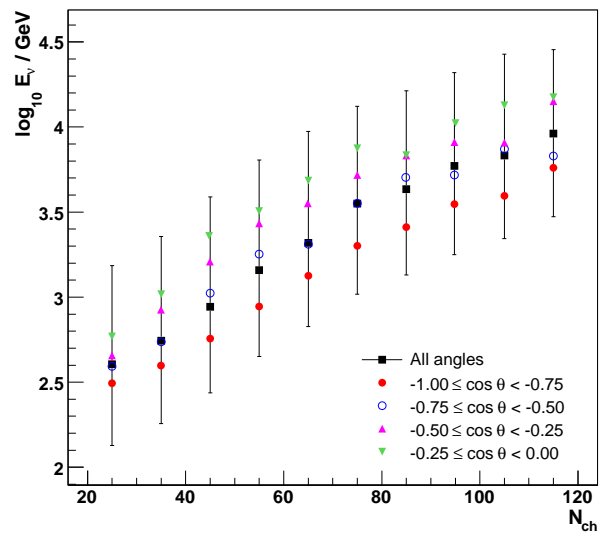}}
\caption{\label{fig:nch_en} Simulated profile histogram of median neutrino
  energy versus number of OMs hit ($\nch$), both for all zenith angles below the
  horizon and for various zenith angle ranges.  Error bars on the all-angle
  points represent the $\pm 1\sigma$ spread at each $\nch$.}  
\end{figure}

\begin{figure*}
\begin{center}
\subfigure{
\includegraphics[scale=0.42]{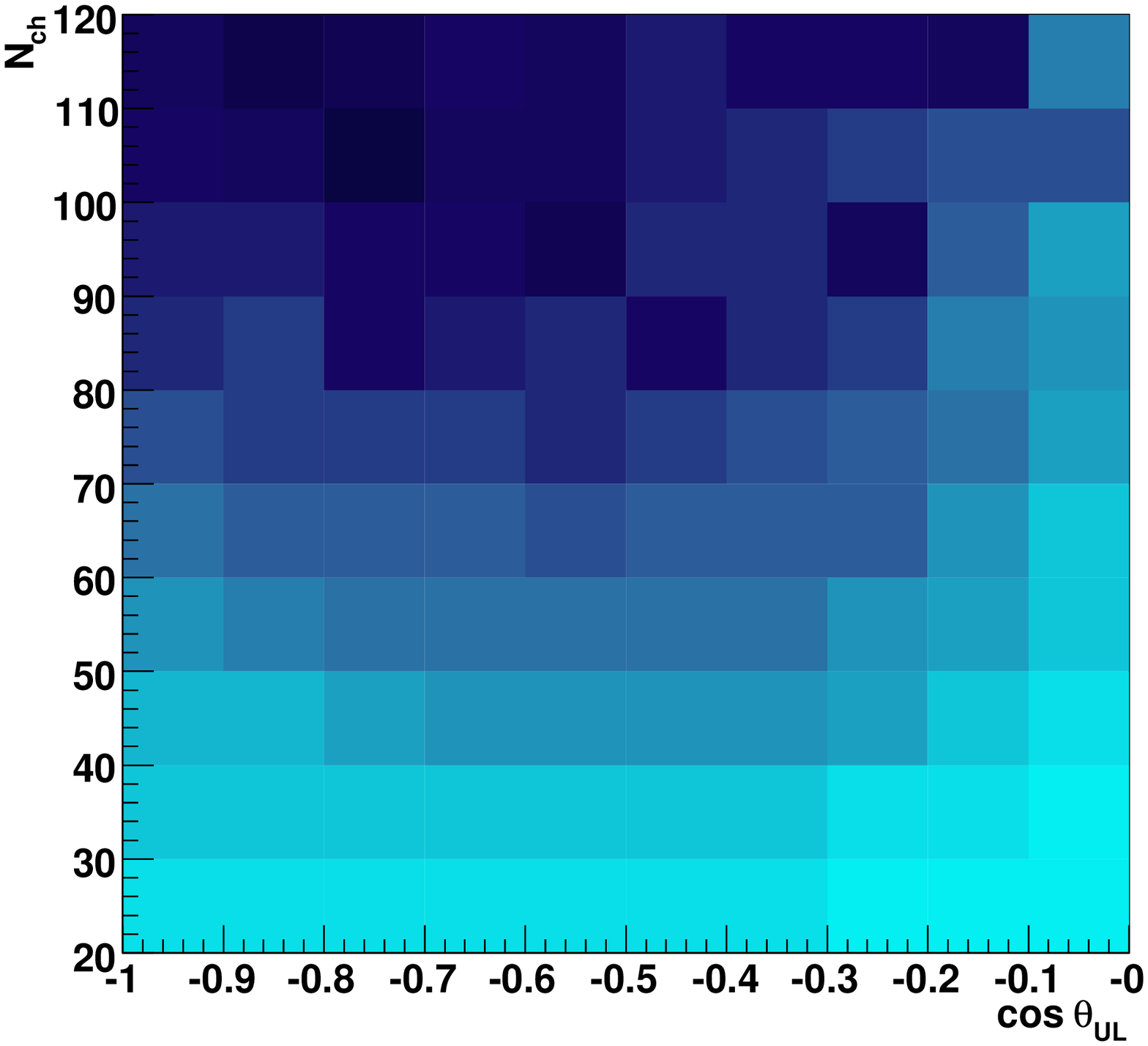}}%
\subfigure{
\includegraphics[scale=0.45]{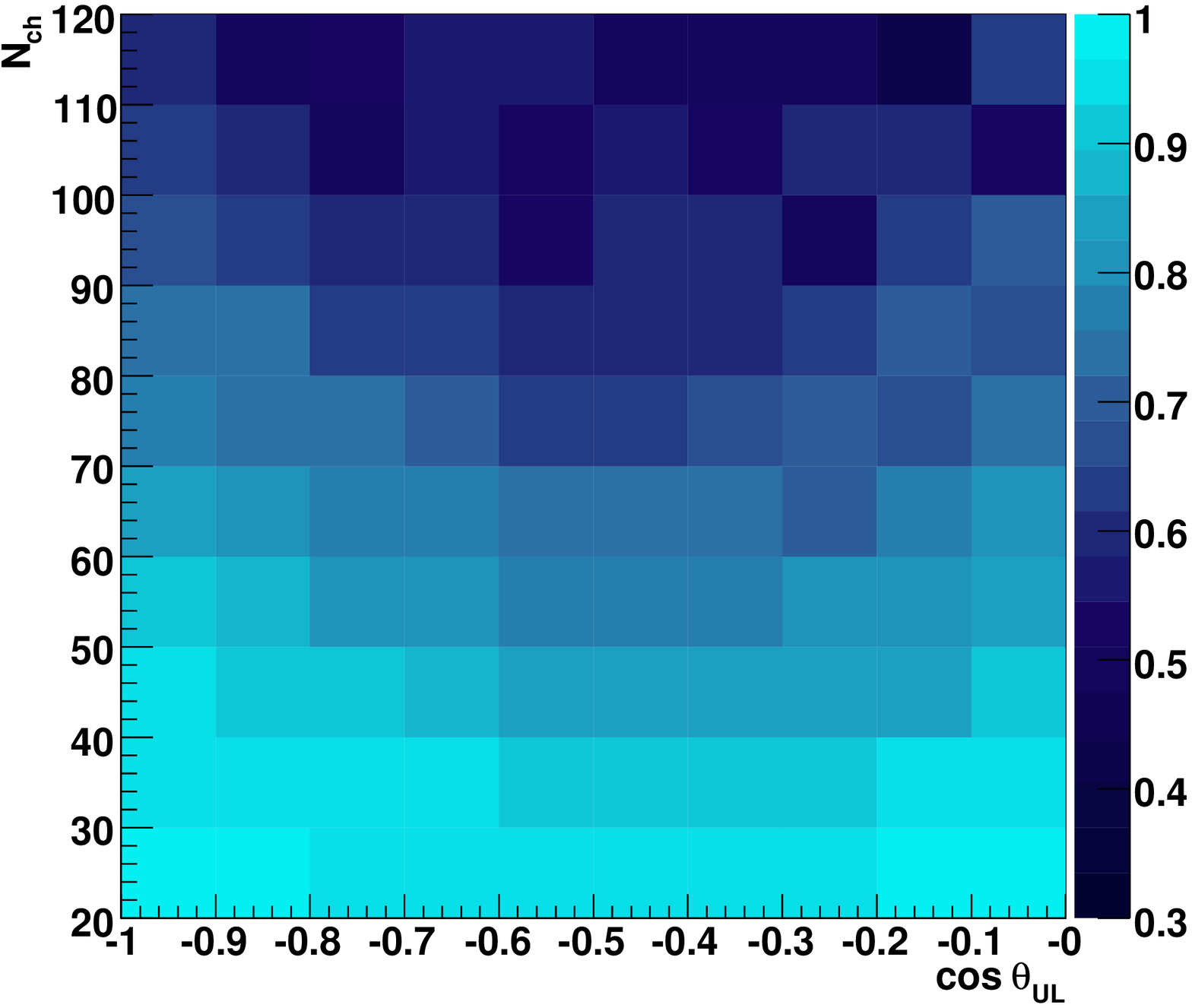}}%
\end{center}
\caption{\label{fig:mc_coszen_nch} Ratio of the simulated number of events given VLI (left, with
  $n=1$, $\sin 2\xi = 1$, and $\Delta\delta = 10^{-26}$) and QD (right, with $n=2$ and $D^* =
  10^{-30}\ \text{GeV}^{-1}$) to conventional oscillation predictions for
  the zenith angle and $\nch$ distribution.}  
\end{figure*}

As described in section \ref{sec:pheno}, the signature of a flavor-changing
new physics effect such as VLI or QD is a deficit of $\nu_{\mu}$ events
at the highest energies and longest baselines (i.e., near the vertical
direction).  For our directional observable, we use the cosine of the
reconstructed zenith angle as given by the UL fit, $\cos \theta_{\text{UL}}$ 
(with $-1$ being the vertical up-going direction).  
We use the number of OMs (or channels) hit, $\nch$, as an energy-correlated
observable.  Fig.~\ref{fig:nch_en} shows the neutrino energy as a function of
the simulated $\nch$ response.  Fig.~\ref{fig:mc_coszen_nch} shows the
simulated effects of QD and VLI on both the zenith angle and $\nch$
distributions, a deficit of events at high $\nch$ and towards more vertical 
directions.  Because the $\nch$ energy estimation is approximate, the VLI
oscillation minima are smeared out, and the two effects look similar in the
observables.  Furthermore, the observable minima are not exactly in the
vertical direction because the $\nch$-energy relationship varies with
zenith angle (see Fig.~\ref{fig:nch_en}), since the detector is taller than
it is wide.  However, this geometry is beneficial for angular
reconstruction of near-vertical events and so is still well-suited to this
analysis.

\subsection{Statistical Methods}

To test the compatibility of our measured atmospheric neutrino $(\cos\theta_{\text{UL}},
N_\text{ch})$ distribution with the various
hypotheses characterized by the VLI and QD parameters, we turn to the
frequentist approach of Feldman and Cousins \cite{fc}.
Specifically, we iterate over our physics parameters ${\theta_r}$, and our
test statistic at each point in the parameter space is the log likelihood ratio
comparing this to the best-fit point ${\hat{\theta}_r}$,

\bea
\Delta\mathcal{L}(\theta_r) & = & \mathcal{L}(\theta_r) -
\mathcal{L}(\hat{\theta}_r) \\
\nonumber & = & -2 \ln P(\{n_i\}|\theta_r) + 2 \ln P(\{n_i\}|\hat{\theta}_r) \\
\nonumber & = & 2 \sum_{i=1}^N \left(\mu_i - \hat\mu_i + n_i \ln
\frac{\hat\mu_i}{\mu_i} \right)  
\eea

\nin for binned distributions of observables with $n_i$ counts in the
$i$th bin, with $\mu_i (\hat\mu_i)$ expected given physics parameters
$\theta_r (\hat\theta_r)$.  For example, in a search for VLI effects, our physics
parameters ${\theta_r}$ are the VLI parameters $\log_{10}\Delta\delta$ and 
$\sin 2\xi$; a binned distribution of simulated $\nch$ and
$\cos\theta_{\text{UL}}$ gives us $\mu_i$ for a particular value of the VLI
parameters; and the distribution of $\nch$ and $\cos\theta_{\text{UL}}$ for
the data gives us $n_i$.

As in Ref.~\cite{fc}, we characterize the
spread in the test statistic $\Delta\mathcal{L}$ expected from
statistical variations by generating 
a number of simulated experiments at each point $\theta_r$.   To define
the allowed region of parameter space at a confidence level (CL) $\alpha$,
we find the critical value  $\Delta\mathcal{L}_{\mathrm{crit}}(\theta_r)$ for which a
fraction $\alpha$ of the experiments at $\theta_r$ satisfy $\Delta\mathcal{L} <
\Delta\mathcal{L}_{\mathrm{crit}}$.  Then our acceptance region at this CL
is the set of parameter space $\{\theta_r\}$ where 
$\Delta\mathcal{L}_{\mathrm{data}}(\theta_r) <
\Delta\mathcal{L}_{\mathrm{crit}} (\theta_r)$. 

The above procedure does not \textit{a priori} incorporate any systematic errors
 (or in statistical terms, \textit{nuisance parameters}).  For a review of
recent approaches to this problem, see \cite{cousins}.  We use
an approximation for the likelihood ratio that, in a sense, uses the
worst-case values for the nuisance parameters $\theta_s$ --- the values
that make the data fit the hypothesis the best at the point $\theta_r$.
In other words, we marginalize over $\theta_s$ in both the numerator and
the denominator of the likelihood ratio:

\be
\label{eq_profilellh}
\Delta\mathcal{L}_p(\theta_r) = \mathcal{L}(\theta_r, \hat{\hat{\theta}}_s) -
\mathcal{L}(\hat{\theta}_r, \hat{\theta}_s)\ ,
\ee

\nin where we have globally minimized the second term, and we have
conditionally minimized the first term, keeping $\theta_r$ fixed but varying
the nuisance parameters to find $\hat{\hat{\theta}}_s$.  This test
statistic is known as the \textit{profile likelihood} \cite{kendall}.  

The profile likelihood is used in combination with a $\chi^2$
approximation in the \textsc{minos} method in \textsc{minuit} \cite{minuit} and is also
explored in some detail by Rolke \emph{et al.}~\cite{rolke2000,rolke2004}.
To extend our frequentist construction to the profile
likelihood, we follow the \textit{profile construction} method
\cite{feldman_pc,cranmer05}: we 
perform simulated experiments as before, but instead of iterating through
the entire $(\theta_r, \theta_s)$ space, at each point in the physics
parameter space $\theta_r$ we fix $\theta_s$ to its best-fit value from the
\emph{data}, $\hat{\hat{\theta}}_s$. Then we recalculate the profile likelihood for
the experiment as defined in Eq.~\ref{eq_profilellh}.  As before, this
gives us a set of likelihood ratios $\{\Delta\mathcal{L}_p\}$ with which we can
define the critical value for a confidence level that depends only on
$\theta_r$.  

\subsection{\label{sec:systematics}Systematic Errors}

Each nuisance parameter added to the likelihood test statistic increases
the dimensionality of the space we must search for the minimum; therefore,
to add systematic errors we group by their effect on the $(\cos\theta_{\text{UL}},
\nch)$ distribution.  We define the following four classes of
errors: 1) \textit{normalization} errors, affecting only the 
total event count; 2) \textit{slope} errors, affecting the energy
spectrum of the neutrino events and thus the $\nch$ distribution;
3) \textit{tilt} errors, affecting the $\cos\theta_{\text{UL}}$ distribution; and 4)
\textit{OM efficiency} errors, which affect the probability of photon
detection and change both the $\cos\theta_{\text{UL}}$ and $\nch$
distribution.  These errors are incorporated into the simulation as
follows:

\begin{itemize}

\item{Normalization errors are incorporated via a uniform weight $1 \pm
  \sqrt{(\alpha_1^2 + \alpha_2^2)}$;}

\item{slope errors are incorporated via an energy-dependent event weight
  $(E/E_{\text{median}})^{\Delta\gamma}$, where $E_{\text{median}}$ is the 
median neutrino energy at the final cut level, 640 GeV;}

\item{tilt errors are incorporated by 
linearly tilting the $\cos\theta_{\text{UL}}$ distribution via a factor 
$1 + 2\kappa (\cos\theta_{\text{UL}} + \frac{1}{2})$;}

\item{and OM efficiency errors are incorporated by regenerating
  atmospheric neutrino simulation while changing the efficiency of all OMs
  in the detector simulation from the nominal value by a factor $1+\epsilon$.}

\end{itemize}

\nin We split the
normalization error into two components, $\alpha_1$ and $\alpha_2$, to
facilitate the determination of the conventional atmospheric flux, as we
discuss later.

Table \ref{tab:systematics} summarizes
sources of systematic error and the class of each error.  The total
normalization errors $\alpha_1$ and $\alpha_2$ are obtained by adding the
individual normalization errors in 
quadrature, while the tilt $\kappa$ and slope change
$\Delta\gamma$ are added linearly.  Asymmetric error totals
are conservatively assumed to be symmetric, using whichever deviation from
the nominal is largest. Each class of error maps to one dimension in the
likelihood space, so for example in the VLI case,  $\mathcal{L}(\theta_r, \theta_s) =
\mathcal{L}(\Delta\delta, \sin 2\xi, \alpha, \Delta\gamma, \kappa,
\epsilon)$.  During minimization, each nuisance 
parameter is allowed to vary freely within the range allowed around its
nominal value, with each point in the likelihood space giving a specific prediction for
the observables, $\nch$ and $\cos\theta_{\text{UL}}$.  In most cases, the
nominal value of a nuisance parameter corresponds to the 
predictions of the Barr \textit{et al.} flux, with best-known inputs to the
detector simulation chain. 

\begin{table}
\caption{\label{tab:systematics} Systematic errors in the atmospheric muon neutrino
  flux, separated by effect on the observables $\cos\theta_{\text{UL}}$ and
  $\nch$ (see section \ref{sec:systematics} for details on the parameters).}
\begin{ruledtabular}
\begin{tabular}{|l|c|c|}
  Error & Class & Magnitude \\
  \hline
  Atm. $\nu_{\mu}+\bar{\nu}_{\mu}$ flux & $\alpha_1$ & $\pm 18\%$ \\
  Neutrino interaction & $\alpha_2$ & $\pm 8\%$ \\
  Reconstruction bias & $\alpha_2$ & $-4\%$ \\
  $\nu_\tau$-induced muons & $\alpha_2$ & $+2\%$ \\
  Background contamination & $\alpha_2$ & $+1\%$ \\
  Charmed meson contribution & $\alpha_2$ & $+1\%$ \\
  Timing residual uncertainty & $\alpha_2$ & $\pm 2\%$ \\
  Muon energy loss & $\alpha_2$ & $\pm 1\%$ \\
  \hline
  Primary CR slope (H, He) & $\Delta\gamma$ & $\pm 0.03$ \\
  Charmed meson contribution & $\Delta\gamma$ & $+0.05$ \\
  \hline
  Pion/kaon ratio & $\kappa$ & $+0.01$/$-0.03$ \\
  Charmed meson contribution & $\kappa$ & $-0.03$ \\
  \hline
  OM efficiency, ice & $\epsilon$ & $\pm 10\%$ \\
\end{tabular}
\end{ruledtabular}
\end{table}

One of the largest sources of systematic error is the overall normalization of the
atmospheric neutrino flux.  While the total $\nu_\mu+\bar\nu_\mu$ simulated
event rate for recent models \cite{bartol,honda} only differs by $\pm7\%$,
this masks significantly larger differences in the individual $\nu_\mu$ and
$\bar\nu_\mu$ rates.  We take the latter difference of $\pm18\%$ to be more
representative of the true uncertainties in the models.  This is also in line
with the total uncertainty in the flux estimated in Ref.~\cite{honda}.

Another large source of error in the event rate arises from
uncertainties in our simulation of the neutrino interactions, including the
neutrino-nucleon cross section, parton distribution functions, and the
neutrino-muon scattering angle.  We quantify this by comparing our
\textsc{nusim} simulation with a sample generated with the \textsc{anis}
simulator \cite{anis}.  \textsc{anis} uses the CTEQ5
cross sections and parton distribution functions \cite{cteq5},
compared to MRS \cite{mrs} in \textsc{nusim}, and it also accurately simulates
the neutrino-muon scattering angle.  We find an 8\% difference in the
normalization for an atmospheric neutrino spectrum.

A third significant source of error is the uncertainty in the
efficiency of the optical modules, that is, the probability an OM will detect
a Cherenkov photon.  This has a large effect on both the overall
detector event rate (a decrease of 1\% in efficiency results in a decrease
of 2.5\% in event rate) and the shape of the zenith angle and $\nch$
distributions.  We quantify the uncertainty by comparing the trigger
rate of down-going muons with simulation predictions given various OM
efficiencies, including the uncertainty of hadronic interactions by using
\textsc{corsika} air shower simulations with the \textsc{sibyll 2.1} \cite{sibyll},
\textsc{epos 1.60} \cite{epos}, and \textsc{qgsjet-ii-03} \cite{qgs2} interaction models.  We
find that we can constrain the optical module efficiency to within
$+10\%$/$-7\%$, consistent with the range of uncertainty determined in
Ref.~\cite{amanda5yr}. Furthermore, because uncertainties in the ice properties have
similar effects on our observables, we model OM efficiency and ice
scattering/absorption together as a single source of error of $\pm10\%$ (in
efficiency).  

Other smaller sources of error were quantified with dedicated simulation
studies or, if directly applicable to this analysis, taken from
Ref.~\cite{amanda5yr}.  For example, we determine the effect of a large
contribution of ``prompt'' $\nu_\mu$ from charmed particle decay by
simulating the optimistic Naumov RQPM flux \cite{naumov}, and find that its
effects can be modeled with the normalization, slope, and tilt
errors as shown in table \ref{tab:systematics}.  Finally, we characterize
our uncertainty in our reconstruction quality parameters (``reconstruction
bias'' in table \ref{tab:systematics}) by investigating how systematic
disagreements between data and simulation affect the number of events
surviving to the final cut level. 

\subsection{{\label{sec:parameters}}Binning and Analysis Parameters}

In general, finer binning provides higher sensitivity with a likelihood
analysis, and indeed we find a monotonic increase in sensitivity to VLI
effects while increasing the number of bins in $\cos\theta_{\text{UL}}$ and
$\nch$.  However, because the further gains in sensitivity are minimal with binning
finer than $10\times10$, we limit ourselves to this size in order to avoid
any systematic artifacts that might show up were we to bin, say, finer than
our angular resolution.  We also limit the $\nch$ range for the
analysis to $20 \leq \nch < 120$.  While the multiplicity trigger requires
24 or more OMs in an event, the hit-cleaning algorithms reduce the
effective threshold to $\nch \approx 20$.  We limit the high energy range to
events with $\nch < 120$ in order to avoid regions with poor
statistics.   This limits the possibility that a few remaining 
background events concentrated at high energy might bias the analysis, which
assumes the data can be modeled by atmospheric neutrino simulation with a
small energy-\textit{independent} background contamination.  The choice of $\nch$
range reduces the number of candidate neutrino events in the analysis region to 5511.
These binning choices were made in a blind
manner, using simulation to determine sensitivity to the new physics effects.

We also make a few more simplifications to reduce the dimensionality of the
likelihood space.  First, the phase $\eta$ in the VLI survival
probability (Eq. \ref{vli_psurv}) is only relevant if the VLI effects are
large enough to overlap in energy with conventional oscillations (i.e.,
below $\sim$100 GeV).  Since our neutrino sample is largely outside this range,
we set $\cos\eta = 0$ for this search.  This means we can also limit the
VLI mixing angle to the range $0 \le \sin 2\xi \le 1$.  Second, in the QD
case, we vary the decoherence parameters $D^*_i$ in 
pairs $(D^*_3,D^*_8)$ and $(D^*_6,D^*_7)$.  If we set $D^*_3$ and 
$D^*_8$ to zero, after decoherence $1/2$ of $\nu_{\mu}$ remain; with
$D^*_6$ and $D^*_7$ set to zero, $5/6$ remain; and with
all $D^*_i$ equal and nonzero, $1/3$ remain after decoherence.
These limiting behaviors are relevant when considering sensitivity to
different parts of the parameter space.  

Finally, in the absence of new physics, we can use the same methodology to
determine the conventional atmospheric neutrino flux. 
In this case, the nuisance parameters $\alpha_1$ (the uncertainty on the
atmospheric neutrino flux normalization) and $\Delta\gamma$ (the change in
spectral slope relative to the input model) become our physics parameters.
The determination of an input energy spectrum by using a set of model
curves with a limited number of parameters is commonly known as
\textit{forward-folding} (see e.g.~Ref.~\cite{forward_folding}).  

Table \ref{tab:params} summarizes the likelihood parameters used 
for the VLI, QD, and conventional analyses. 

\begin{table}
\caption{\label{tab:params} Physics parameters and nuisance parameters used
  in each of the likelihood analyses (VLI, QD, and conventional).}
\begin{ruledtabular}
\begin{tabular}{|l|c|c|}
  Analysis & Physics parameters & Nuisance parameters \\
  \hline
  VLI & $\Delta\delta,\ \sin 2\xi$ &
  $\alpha_1,\ \alpha_2,\ \Delta\gamma,\ \kappa,\ \epsilon$ \\
  QD & $D^*_{3,8},\ D^*_{6,7}$ &
  $\alpha_1,\ \alpha_2,\ \Delta\gamma,\ \kappa,\ \epsilon$ \\
  Conv. & $\alpha_1,\ \Delta\gamma$ &
  $\alpha_2,\ \kappa,\ \epsilon$ \\
\end{tabular}
\end{ruledtabular}
\end{table}


\section{\label{sec:results}Results}

\begin{figure*}
\begin{center}
\subfigure{
\includegraphics[scale=0.45]{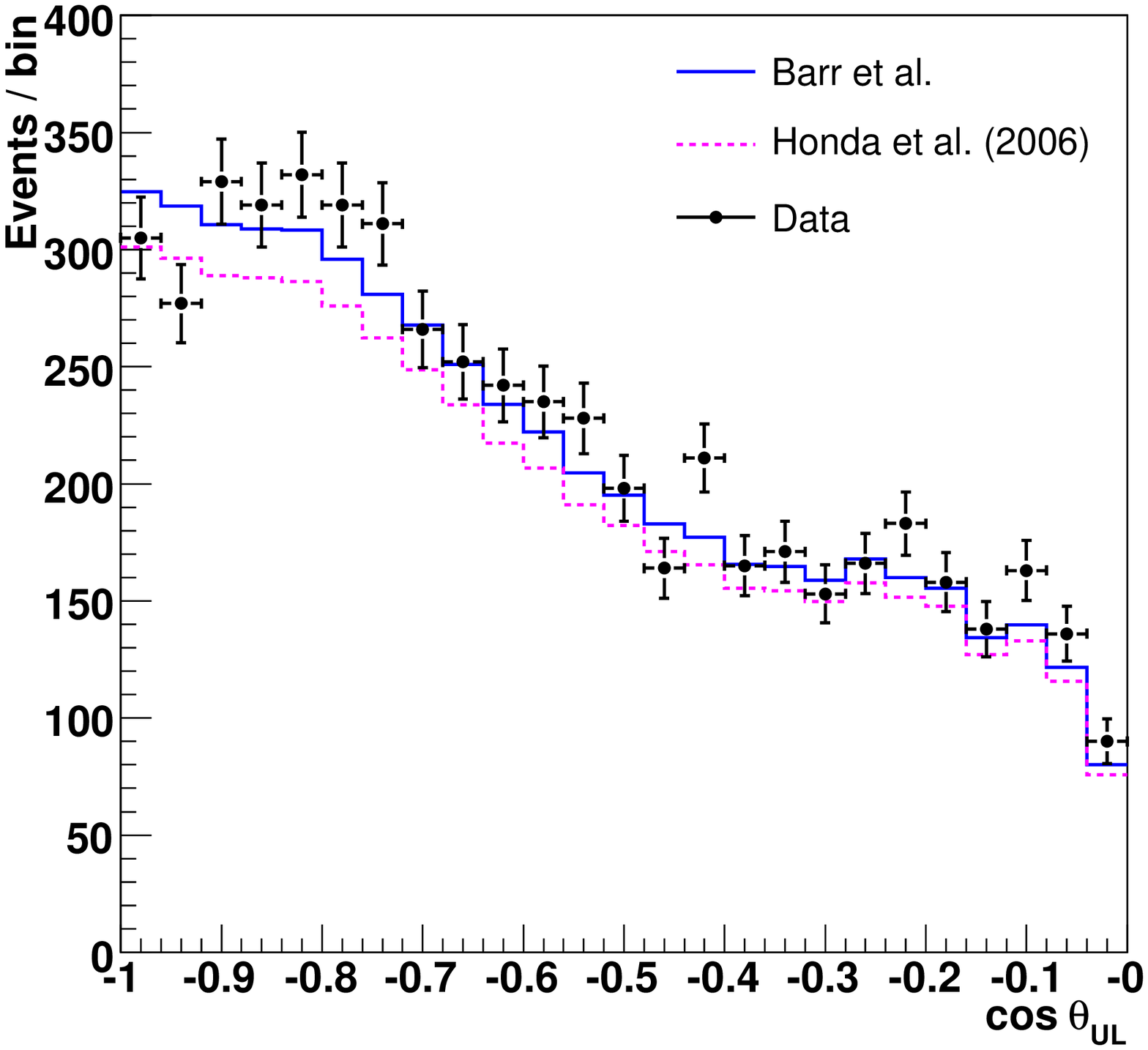}}%
\subfigure{
\includegraphics[scale=0.45]{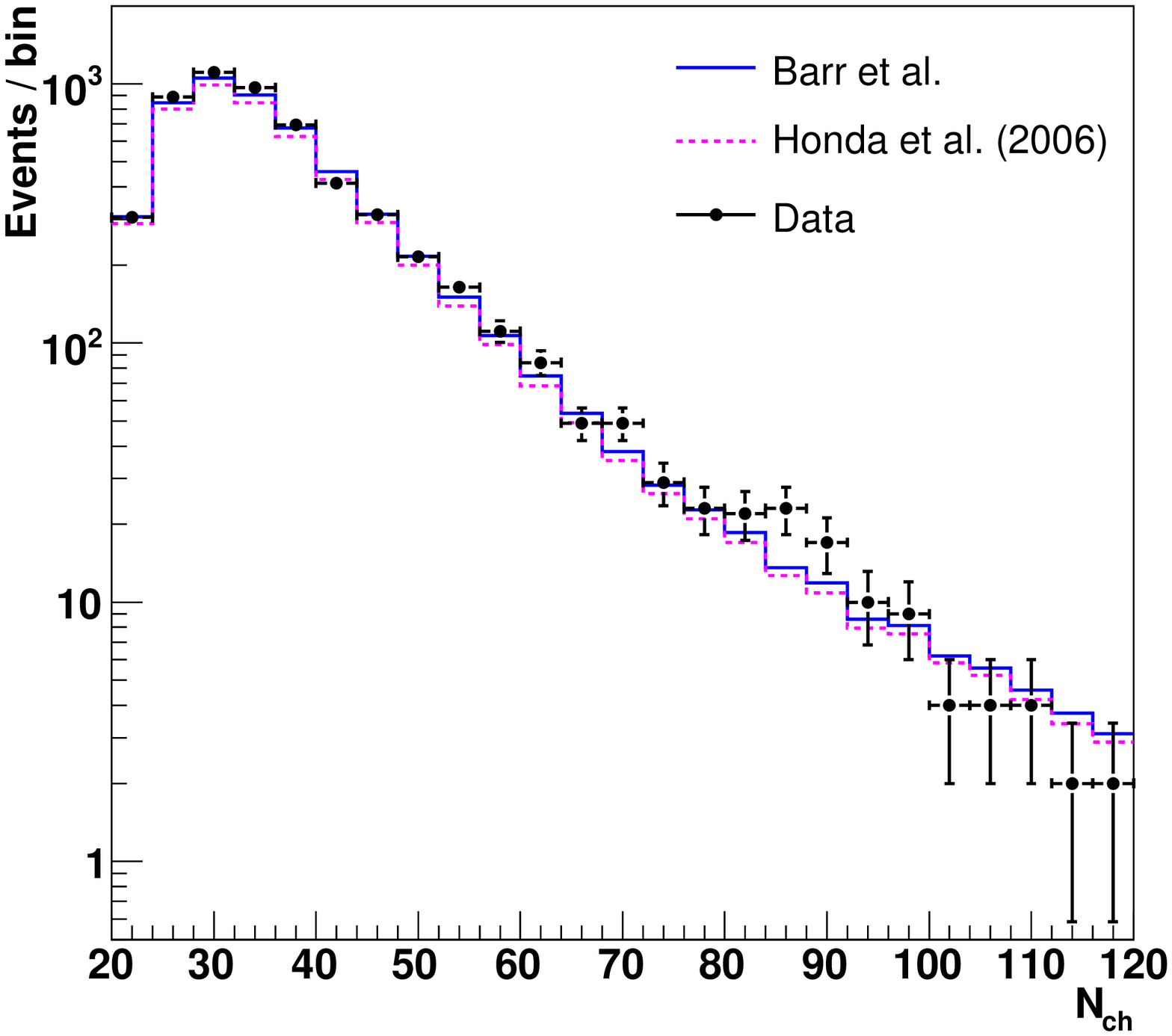}}%
\end{center}
\caption{\label{fig:data_coszen_nch} Zenith angle and $\nch$ distribution
  of candidate atmospheric neutrino events in the final sample, compared with Barr
  \textit{et al.}~\cite{bartol} and Honda \textit{et al.}~\cite{honda}
  predictions (statistical error bars).}
\end{figure*}

After performing the likelihood analysis on the $(\cos\theta_{\text{UL}},
\nch)$ distribution, we find no evidence for VLI-induced oscillations or
quantum decoherence, and the data are consistent with expectations from
atmospheric flux models.  The reconstructed zenith angle and $\nch$
distributions compared to standard atmospheric neutrino models are shown
in Fig.~\ref{fig:data_coszen_nch}, projected into one dimension from the $10\times10$
two-dimensional analysis distribution and rebinned.  Given the lack of evidence
for new physics, we set upper limits 
on the VLI and QD parameters. 

\subsection{Upper Limits on Violation of Lorentz Invariance}

\begin{figure}
\resizebox{0.45\textwidth}{!}{\includegraphics{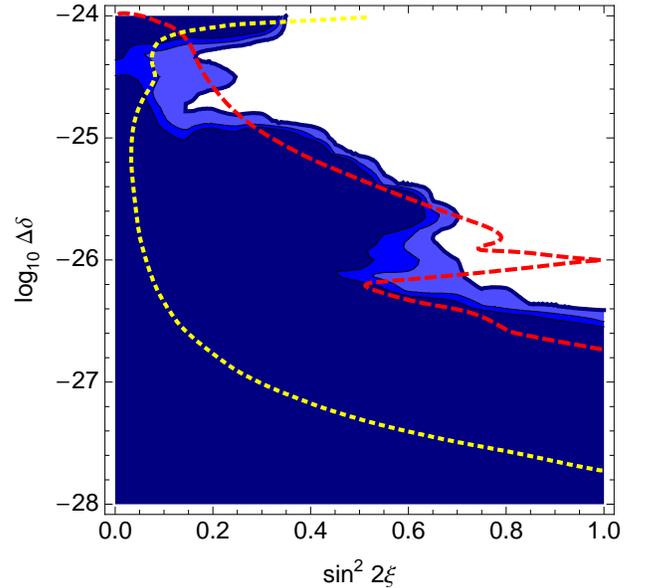}}
\caption{\label{fig:vli_e1_limit} 90\%, 95\%, and 99\% CL allowed regions
  (from darkest to lightest) for VLI-induced oscillation effects with
  $n=1$.  Note we plot $\sin^2 2\xi$ to enhance the region
  of interest.  Also shown are the Super-Kamiokande + K2K 90\% contour
  \cite{ggm04} (dashed line), and the projected IceCube 10-year 90\% sensitivity
  \cite{icecube_vli} (dotted line).}
\end{figure}

The 90\% CL upper limits on the VLI parameter
$\Delta\delta$ for oscillations of 
various energy dependencies, with maximal mixing ($\sin 2\xi = 1$) and phase
$\cos\eta = 0$, are presented in table \ref{tab:vli_qd_limits}.  Allowed
regions at 90\%, 95\%, and 99\% confidence levels in the
$\Delta\delta\text{-}\sin 2\xi$ plane for the $n=1$ hypothesis are shown in
Fig.~\ref{fig:vli_e1_limit}.  The upper limit at maximal mixing of $\Delta\delta
\le 2.8\times10^{-27}$ is competitive with that from a combined
Super-Kamiokande and K2K analysis \cite{ggm04}. 

In the $n=1$ case, recall that the VLI parameter $\Delta\delta$ corresponds
to the splitting in velocity eigenstates $\Delta c/c$.  Observations of
ultra-high energy cosmic rays constrain VLI velocity splitting in other
particle sectors, with the upper limit on proton-photon splitting of $(c_p-c)/c <
10^{-23}$ \cite{coleman99}.  While we probe a rather specific manifestation
of VLI in the neutrino sector, our limits are orders of magnitude better
than those obtained with other tests.  

\subsection{Upper Limits on Quantum Decoherence}

\begin{figure}
\resizebox{0.45\textwidth}{!}{\includegraphics{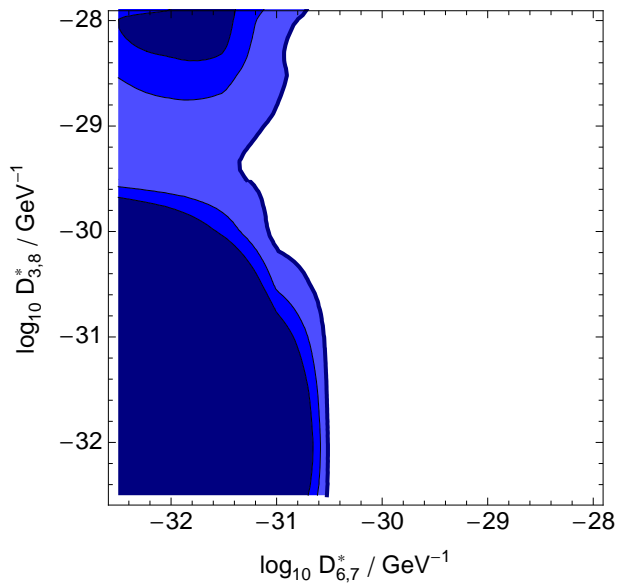}}
\caption{\label{fig:qd_e2_limit} 90\%, 95\%, and 99\% CL allowed regions
  (from darkest to lightest) for QD effects with $n=2$.}
\end{figure}

The 90\% CL upper limits on the decoherence parameters $D_i^*$ given various
energy dependencies are also shown in table \ref{tab:vli_qd_limits}.  Allowed
regions at 90\%, 95\%, and 99\% confidence levels in the
$D^*_{3,8}\text{-}D^*_{6,7}$ plane for the $n=2$ case are shown in
Fig.~\ref{fig:qd_e2_limit}.  The 90\% CL upper limit from this analysis with all $D^*_i$
equal for the $n=2$ case, $D^* \le 1.3\times10^{-31}\ \text{GeV}^{-1}$,
extends the previous best limit from Super-Kamiokande by nearly four orders
of magnitude.  Because of the strong $E^2$ energy dependence, AMANDA-II's extended
energy reach allows much improved limits.  

\begin{table}
\caption{\label{tab:vli_qd_limits} 90\% CL upper limits from this analysis on VLI
  and QD effects proportional to $E^n$.  VLI upper limits are for the case of
  maximal mixing ($\sin 2\xi = 1$), and QD upper limits are for the case of
  $D^*_3 = D^*_8 = D^*_6 = D^*_7$.} 
\renewcommand\arraystretch{1.2}
\begin{tabular}{c|c|c|c}
  $n$ & VLI ($\Delta\delta$) & QD ($D^*$) & Units \\
  \hline
  1 & $\ 2.8\times10^{-27}\ $ & $\ 1.2\times10^{-27}\ $ & -- \\ 
  2 & $\ 2.7\times10^{-31}\ $ & $\ 1.3\times10^{-31}\ $ & $\ \text{GeV}^{-1}$ \\
  3 & $\ 1.9\times10^{-35}\ $ & $\ 6.3\times10^{-36}\ $ & $\ \text{GeV}^{-2}$ \\
\end{tabular}
\end{table}

\subsection{Determination of Atmospheric Flux}

In the absence of evidence for violation of Lorentz invariance or quantum
decoherence, we interpret the atmospheric neutrino flux in the context of
Standard Model physics only.  We use the likelihood analysis to perform a
two-parameter forward-folding of the atmospheric neutrino flux to determine
the normalization and any change in spectral index relative to existing
models.  As described in section \ref{sec:parameters}, we test hypotheses of the form 

\be
\frac{d\Phi}{dE} = (1+\alpha_1)\ \frac{d\Phi_{\text{ref}}}{dE}
\left(\frac{E}{E_{\text{median}}}\right)^{\Delta\gamma} \ ,  
\ee

\nin where $d\Phi_{\text{ref}}/dE$ is the differential Barr
\textit{et al.} or Honda \textit{et al.} flux.  

\begin{figure}
\resizebox{0.45\textwidth}{!}{\includegraphics{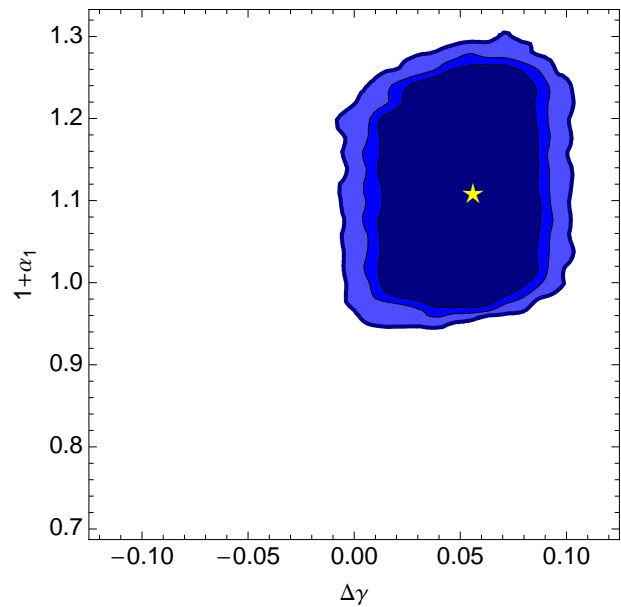}}
\caption{\label{fig:conv_limit} 90\%, 95\%, and 99\% allowed regions
  (from darkest to lightest) for the normalization ($1+\alpha_1$) and change in spectral
  index ($\Delta\gamma$) of the conventional atmospheric neutrino flux, relative to Barr
  \textit{et al.} \cite{bartol}.  The star marks the central best-fit
  point.}
\end{figure}

\begin{figure}
\resizebox{0.45\textwidth}{!}{\includegraphics{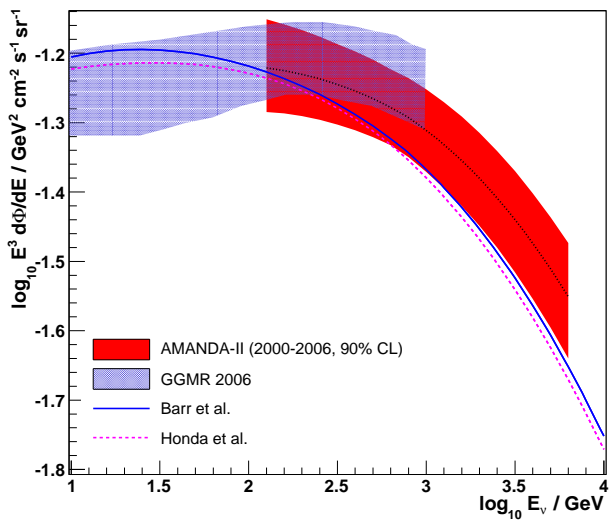}}
\caption{\label{fig:flux_band} Angle-averaged 
  $\nu_{\mu}+\bar\nu_\mu$ atmospheric neutrino flux (solid band, 90\% CL
  from the forward-folding analysis), multiplied by $E^3$ to enhance
  features.  The dotted line shows the central best-fit curve.  Also shown is
  a previous result by Gonz\'alez-Garc\'ia \textit{et al.}~using
  Super-Kamiokande data \cite{maltoni_atm}, as well 
  as Barr \textit{et al.} \cite{bartol} and Honda \textit{et al.} \cite{honda}
  predictions.  All fluxes are shown prior to oscillations.}  
\end{figure}

The allowed regions in the $\alpha_1$-$\Delta\gamma$ parameter space are
shown in Fig.~\ref{fig:conv_limit}.  We display the band of allowed
energy spectra in Fig.~\ref{fig:flux_band}, where we have constructed the
allowed region by forming the envelope of the set of curves allowed on the
90\% contour in Fig.~\ref{fig:conv_limit}.  The energy range of the band is
the intersection of the 5\%-95\% regions of the allowed set of spectra, so
restricted in order to limit the range of our constraints to an energy
region in which AMANDA-II is sensitive.  

The central best-fit point is also shown in Figs.~\ref{fig:conv_limit} and
\ref{fig:flux_band}.  In fact, there is actually a range of best-fit points
for the normalization, because of the degeneracy between the normalization
parameter $\alpha_1$ and the systematic error $\alpha_2$.  Specifically, we
find the best-fit spectra to be 

\be
\frac{d\Phi_{\text{best-fit}}}{dE} = (1.1\pm0.1)
\left(\frac{E}{640\ \text{GeV}}\right)^{0.056}\cdot\frac{d\Phi_{\text{Barr}}}{dE}
\ee

\nin for the energy range 120 GeV to 7.8 TeV, where the $\pm0.1$ is not the
error on the fit but the range of possible best-fit values.  This result is
compatible 
with an analysis of Super-Kamiokande data 
\cite{maltoni_atm} as well as an unfolding of the Fr\'ejus data
\cite{frejus}, and extends the Super-Kamiokande measurement by nearly an
order of magnitude in energy.  Our data suggest an atmospheric neutrino
spectrum with a slightly harder spectral slope and higher normalization
than either the Barr \textit{et al.} or Honda \textit{et al.} model.
The likelihood ratio $\Delta\mathcal{L}$ of the unmodified Barr
\textit{et al.} spectrum (at the point (0,1) in Fig.~\ref{fig:conv_limit})
to the best-fit point is 
4.9, corresponding to the 98\% CL. 

\subsection{Discussion and Future Prospects}

To summarize, we have set stringent upper limits on both Lorentz violation
and quantum decoherence effects
in the neutrino sector, with a VLI upper limit at the 90\% CL of $\Delta\delta = \Delta c/c <
2.8 \times 10^{-27}$ for VLI oscillations proportional to the neutrino
energy $E$, and a
QD upper limit at the 90\% CL of $D^* < 1.3 \times 10^{-31} \text{GeV}^{-1}$ for
decoherence effects proportional to $E^2$.  We have also set upper limits
on VLI and QD effects with different energy dependencies.  Finally, we have
determined the atmospheric neutrino spectrum in the energy range from 120
GeV to 7.8 TeV and find a best-fit result that is slightly higher in
normalization and has a harder spectral slope than either the Barr \textit{et al.}~or
Honda \textit{et al.}~model.  This result is consistent with
Super-Kamiokande data and extends that measurement by nearly an order of
magnitude in energy.

For an interpretation of the VLI and QD upper limits, we consider natural
expectations for the values of such parameters.  Given 
effects proportional to $E^2$ and $E^3$, one can argue via dimensional 
analysis that the new physics parameter should contain a
power of the Planck mass $M_{\text{Pl}}$ or $M_{\text{Pl}}^2$,
respectively \cite{cygnus_ob2}.  For example, for the decoherence
parameters $D$, we may expect 

\bea
D & = & D^*E_\nu^{n}\nonumber\\
 & = & d^*\frac{E_\nu^{n}}{M_{\text{Pl}}^{n-1}}
\eea

\nin for $n \ge 2$, and $d^*$ is a dimensionless quantity that is $O(1)$ by
naturalness.  From the limits in table \ref{tab:vli_qd_limits} we find $d^*
< 1.6\times10^{-12}\ (n=2)$ and $d^* < 910\ (n=3)$.  For the $n=2$ case, the
decoherence parameter is far below the natural expectation, suggesting
either a stronger suppression than described, or that we have indeed probed
beyond the Planck scale and found no decoherence of this type.

While the AMANDA-II data acquisition system used in this analysis ceased
taking data at the end of 2006, the next-generation, cubic-kilometer-scale
IceCube detector has 
the potential to improve greatly upon the limits presented here, as increased
statistics of atmospheric neutrinos at the highest energies probe smaller
deviations from the Standard Model.
In particular, IceCube should be sensitive to
$n=1$ VLI effects an order of magnitude smaller than the limits from this
analysis \cite{icecube_vli} (see also Fig.~\ref{fig:vli_e1_limit}).  
We also note that we have also only tested one particular manifestation of VLI in the
neutrino sector.  A search of the atmospheric neutrino data for an
unexpected directional dependence (for example, in right ascension) could
probe other VLI effects, such as a universal directional asymmetry (see
e.g.~\cite{mewes}).

Moving beyond searches with atmospheric neutrinos, once high energy astrophysical
neutrinos are detected, analysis of the flavor ratio at Earth can probe
VLI, QD, and CPT violation \cite{cygnus_ob2,hooper_cpt}. Another technique
is to probe VLI via the potential time delays between photons and neutrinos
from gamma-ray bursts (GRBs).  Given the cosmological
distances traversed, this delay could range from $1\ \mu\text{s}$ to 1 year,
depending on the power of suppression by $M_{\text{Pl}}$
\cite{amelino-camelia_grb}.  Detection of high energy neutrinos from multiple
GRBs at different redshifts would allow either confirmation of the delay
hypothesis or allow limits below current levels by several orders of
magnitude \cite{piran}.  Such a search is complicated by the low expected
flux levels from individual GRBs, as well as uncertainty of any intrinsic
$\gamma-\nu$ delay due to production mechanisms in the source (for a
further discussion, see Ref.~\cite{halzen_grb}).  Other probes of Planck-scale
physics may be possible as well, but ultimately this will depend on the
characteristics of the neutrino sources detected.


\begin{acknowledgments}

We acknowledge the support of the following agencies: the U.S.~National
Science Foundation -- Office of Polar Programs; the U.S.~National Science
Foundation -- Physics Division; the University of Wisconsin Alumni Research
Foundation; the U.S.~Department of Energy and National Energy Research
Scientific Computing Center; the Louisiana Optical Network Initiative
(LONI) grid computing resources; the Swedish Research Council; the Swedish Polar
Research Secretariat; the Knut and Alice Wallenberg Foundation, Sweden;
the German Ministry for Education and Research (BMBF); the Deutsche
Forschungsgemeinschaft (DFG), Germany; the Fund for Scientific Research
(FNRS-FWO), Flanders Institute to encourage scientific and technological
research in industry (IWT); the Belgian Federal Science Policy Office (Belspo);
and the Netherlands Organisation for Scientific Research (NWO).  M.~Ribordy
acknowledges the support of the SNF (Switzerland), and A.~Kappes and A.~Gro{\ss}
acknowledge the support of the EU Marie Curie OIF Program. 

\end{acknowledgments}


\appendix*

\section{Formalism}

We present for the interested reader more detail of the phenomenological
background to the atmospheric $\nu_\mu$ survival probabilities for
the VLI and QD hypotheses that we test in this work.  

\subsection{Violation of Lorentz Invariance}

The Standard Model Extension (SME) provides an effective field-theoretic
approach to violation of Lorentz invariance (VLI) \cite{sme}.  The ``minimal''
SME adds all coordinate-independent renormalizable 
Lorentz- and CPT-violating terms to the Standard Model Lagrangian.  Even when
restricted to first order effects in the neutrino sector, the SME results
in numerous potentially observable effects \cite{mewes, mewes2, tandem}.
To specify one particular model that leads to alternative oscillations at
high energy, we consider only the Lorentz-violating Lagrangian
term 

\be
\frac{1}{2}i(c_L)_{\mu\nu ab}\overline{L}_a\gamma^{\mu}\overleftrightarrow{D}^{\nu}L_b 
\ee

\noindent with the VLI parametrized by the dimensionless coefficient $c_L$
\cite{mewes2}. $L_a$ and $L_b$ are left-handed neutrino doublets with indices running
over the generations $e$, $\mu$, and $\tau$, and $D^\nu$ is the covariant
derivative with $A\overleftrightarrow{D}^\nu B \equiv AD^\nu B - (D^\nu
A)B$.  

We restrict ourselves to rotationally invariant scenarios
with only nonzero time components in $c_L$, and we consider only a
two-flavor system.  The eigenstates of the 
resulting $2\times2$ matrix $c_L^{TT}$ correspond to differing maximal
attainable velocity (MAV) eigenstates.  These may be distinct from either
the flavor or mass eigenstates.  Any difference $\Delta c$ in the
eigenvalues will result in neutrino oscillations.  The above construction
is equivalent to a modified dispersion relationship of the form

\be
E^2 = p^2 c_a^2 + m^2 c_a^4
\ee

\noindent where $c_a$ is the MAV for a particular eigenstate, and in
general $c_a \ne c$ \cite{coleman99, glashow04}.  Given that the mass is
negligible, the energy difference between two MAV eigenstates is equal to
the VLI parameter $\Delta c/c = (c_{a1}-c_{a2})/c$, where $c$ is the
canonical speed of light.  

The effective Hamiltonian $H_\pm$ representing the energy shifts
from both mass-induced and VLI oscillations can be written \cite{ggm04}

\be
\label{vli_hamiltonian}
H_\pm = 
\frac{\Delta m^2}{4E} 
\mathbf{U}_{\theta} \begin{pmatrix}-1& 0\\\hphantom{-}0& 1\end{pmatrix} \mathbf{U}_{\theta}^\dagger +
\frac{\Delta c}{c}\frac{E}{2} 
\mathbf{U}_{\xi} \begin{pmatrix}-1& 0\\\hphantom{-}0& 1\end{pmatrix} \mathbf{U}_{\xi}^\dagger
\ee

\noindent with two mixing angles $\theta$ and $\xi$.  The associated
$2 \times 2$ mixing matrices are 

\be
U_\theta = \begin{pmatrix} \hphantom{-}\cos\theta & \sin\theta
  \\ -\sin\theta & \cos\theta \end{pmatrix}
\ee

\nin and

\be
U_\xi = \begin{pmatrix} \cos\xi & \sin\xi e^{\pm i \eta}
  \\ -\sin\xi e^{\mp i \eta} & \cos\xi \end{pmatrix}
\ee

\nin with $\eta$ representing their relative phase.  Solving the 
Louiville equation for time evolution of the state density matrix $\rho$,

\be
\dot{\rho} = -i[H_\pm,\rho] 
\ee

\nin results in the $\nu_\mu$ survival probability in Eq.~\ref{vli_psurv}.
We refer the reader to Ref.~\cite{ggm04} for more detail.

\subsection{Quantum Decoherence}

Several constructions exist of a phenomenological framework for quantum
decoherence effects \cite{mavromatos_rev}.  A
common approach is to modify the time-evolution of the density matrix
$\rho$ with a dissipative term $\slashed{\delta} H \rho$:

\be
\label{eq_density}
\dot{\rho} = -i[H,\rho] + \slashed{\delta} H \rho\ .
\ee

\noindent One method to model such an open system is via
the technique of Lindblad quantum dynamical semigroups \cite{lindblad}.
Here we outline the 
approach in Ref.~\cite{gago}, to which we refer the reader for more
detail.  In this case we have a set of self-adjoint environmental operators
$A_j$, and Eq.~\ref{eq_density} becomes

\be
\label{eq_lindblad}
\dot{\rho} = -i[H,\rho] + \frac{1}{2} \sum_j ([A_j, \rho A_j] +
[A_j \rho, A_j])\ .
\ee

\nin The hermiticity of the $A_j$ ensures the monotonic increase of
entropy, and in general, pure states will now evolve to mixed states.
The irreversibility of this process implies CPT violation
\cite{mavromatos_rev}.

To obtain specific predictions for the neutrino sector, there are again
several approaches for both two-flavor systems \cite{benatti,morgan_qd} and
three-flavor systems \cite{gago, barenboim_3flav}.  Again, we follow the approach
in \cite{gago} for a three-flavor neutrino system including both
decoherence and mass-induced oscillations.  The dissipative term in
Eq.\ \ref{eq_lindblad} is expanded in the Gell-Mann basis ${F_{\mu}, \mu \in
  \{0,\ldots ,8\}}$, such that 

\be
\frac{1}{2} \sum_j ([A_j, \rho A_j] + [A_j \rho, A_j]) = 
\sum_{\mu,\nu} L_{\mu\nu} \rho_{\mu} F_{\nu}\ .
\ee

\nin At this stage we must choose a form for the decoherence matrix
$L_{\mu\nu}$, and we select the weak-coupling
limit in which $L$ is diagonal, with $L_{00} = 0$ and $L_{ii} =
-D_i, i \in \{1,\ldots ,8\}$.  The $D_i$ are in energy units, and their
inverses represent the characteristic length 
scale(s) over which decoherence effects occur.  Solving this system for atmospheric neutrinos
(where we neglect mass-induced oscillations other than $\nu_{\mu}
\rightarrow \nu_{\tau}$) results in the $\nu_{\mu}$ survival probability given in
Eq.~\ref{qd_psurv}.  

In Eq.~\ref{qd_psurv}, we must impose the condition $\Delta m^2/E >
|D_6-D_7|$, but this is not an issue in the parameter space we explore in
this analysis.  If one wishes to ensure strong conditions such as complete
positivity \cite{benatti}, there may be other inequalities that must be
imposed (see e.g.~the discussion in Ref.~\cite{barenboim_3flav}).  

\hyphenation{PHYSTAT}


\begin{thebibliography}{99}

\bibitem{gambini} R.~Gambini and J.~Pullin, Phys.~Rev.~D {\bf59}, 124021
  (1999). 

\bibitem{madore} J.~Madore, S.~Schraml, P.~Schupp, and J.~Wess,
  Eur.~Phys.~J.~C {\bf16}, 161 (2000).  

\bibitem{samuel} V.~A.~Kosteleck\'y and S.~Samuel, Phys.~Rev.~D {\bf39},
  683 (1989).

\bibitem{amelino05} G.~Amelino-Camelia, C.~L\"ammerzahl, A.~Macias, and
  H.~M\"uller, Gravitation and Cosmology: AIP Conf.~Proc.~{\bf758}, 30
  (2005). 

\bibitem{mattingly05} D.~Mattingly, Living Rev.~Relativity {\bf8}, 5 (2005). 

\bibitem{coleman99} S.~Coleman and S.~L.~Glashow, Phys.~Rev.~D {\bf59}, 116008 (1999).  

\bibitem{glashow04} S.~L.~Glashow, arXiv:hep-ph/0407087.  

\bibitem{hawking} S.~W.~Hawking, Commun.~Math.~Phys.~{\bf87}, 395 (1982).

\bibitem{baikal} V.~A.~Balkanov {\em et al.}, Astropart.~Phys.~{\bf12}, 75
  (1999).

\bibitem{amanda_nature} E.~Andr\'{e}s {\em et al.}, Nature {\bf410}, 441 (2001).

\bibitem{antares} J.~A.~Aguilar {\em et al.}, Astropart.~Phys.~{\bf26}, 314
  (2006). 

\bibitem{ahrens04} J.~Ahrens {\em et al.}, Astropart.~Phys.~{\bf20}, 507 (2004).

\bibitem{honda} M.~Honda, T.~Kajita, K.~Kasahara, S.~Midorikawa, and
T.~Sanuki, Phys.~Rev.~D {\bf75}, 043006 (2007). 

\bibitem{barr_unc} G.~D.~Barr, S.~Robbins, T.~K.~Gaisser, and T.~Stanev,
Phys.~Rev.~D {\bf74}, 094009 (2006).

\bibitem{superk04} Y.~Ashie {\em et al.}, Phys.~Rev.~Lett.~{\bfseries 93}, 101801 (2004). 

\bibitem{soudan03} M.~Sanchez {\em et al.}, Phys.~Rev.~D {\bf68}, 113004 (2003).

\bibitem{macro02} M.~Ambrisio {\em et al.}, Phys.~Lett.~B {\bf566}, 35 (2003).

\bibitem{global_osc_08} G.~L.~Fogli {\em et al.}, Phys.~Rev.~D {\bf78},
  033010 (2008).

\bibitem{ggm04} M.~C.~Gonz\'{a}lez-Garc\'{i}a and M.~Maltoni, Phys.~Rev.~D
  {\bf70}, 033010 (2004).

\bibitem{global_osc} M.~Maltoni, T.~Schwetz, M.~T\'{o}rtola, and
  J.~W.~F.~Valle, New Jour.~of Phys.~{\bf6}, 122 (2004).

\bibitem{macro_vli} G.~Battistoni {\em et al.}, Phys.~Lett.~B {\bf615}, 14 (2005).

\bibitem{superk_vli} G.~L.~Fogli, E.~Lisi, A.~Marrone, and G.~Scioscia,
  Phys.~Rev.~D {\bf60}, 053006 (1999).

\bibitem{k2k} M.~H.~Ahn {\em et al.}, Phys.~Rev.~Lett.~{\bf90}, 041801
  (2003).  

\bibitem{amanda_vli_icrc} J.~Ahrens and J.~L.~Kelley \textit{et al.},
  in Proc.~of 30th ICRC (M\'erida), 2007; arXiv:0711.0353.

\bibitem{antares_vli} D.~Morgan, E.~Winstanley, J.~Brunner, and
  L.~F.~Thompson, Astropart.~Phys.~{\bf29}, 345 (2008). 

\bibitem{alfaro_le2} J.~Alfaro, H.~A.~Morales-T\'{e}cotl, and L.~F.~Urrutia,
  Phys.~Rev.~Lett.~{\bf84}, 2318 (2000).

\bibitem{brustein_le2} R.~Brustein, D.~Eichler, and S.~Foffa, Phys.~Rev.~D
  {\bf65}, 105006 (2002).

\bibitem{gasperini_vep} M.~Gasperini, Phys.~Rev.~D {\bf38}, 2635 (1988).

\bibitem{halprin_vep} A.~Halprin and C.~N.~Leung, Phys.~Rev.~Lett.~{\bf67},
  1833 (1991).

\bibitem{adunas_vep} G.~Z.~Adunas, E.~Rodriquez-Milla, and D.~V.~Ahluwalia,
  Phys.~Lett.~B {\bf485}, 215 (2000).

\bibitem{ellis84} J.~Ellis, J.~S.~Hagelin, D.~V.~Nanopoulos, and
  M.~Srednicki, Nucl.~Phys.~B {\bf241}, 381 (1984). 

\bibitem{gago} A.~M.~Gago, E.~M.~Santos, W.~J.~C.~Teves, and R.~Zukanovich
  Funchal, arXiv:hep-ph/0208166.

\bibitem{barenboim_3flav} G.~Barenboim, N.~E.~Mavromatos, S.~Sarkar, and
  A.~Waldron-Lauda, Nucl.~Phys.~B {\bf758}, 90 (2006).  

\bibitem{dbrane} J.~Ellis, N.~E.~Mavromatos, and D.~V.~Nanopoulos,
  Mod.~Phys.~Lett.~A {\bf12}, 1759 (1997); J.~Ellis, N.~E.~Mavromatos,
  D.~V.~Nanopoulos, and E.~Winstanley, Mod.~Phys.~Lett.~A {\bf12}, 243 (1997). 

\bibitem{superk_qd} E.~Lisi, A.~Marrone, and D.~Montanino,
  Phys.~Rev.~Lett.~{\bf85}, 1166 (2000). 

\bibitem{morgan_qd} D.~Morgan, E.~Winstanley, J.~Brunner, and L.~Thompson,
  Astropart.~Phys.~{\bf25}, 311 (2006).

\bibitem{superk_k2k_qd} G.~L.~Fogli, E.~Lisi, A.~Marrone, and D.~Montanino,
  Phys.~Rev.~D {\bf67}, 093006 (2003).

\bibitem{kamland_qd} G.~L.~Fogli, E.~Lisi, A.~Marrone, D.~Montanino, and
A.~Palazzo, Phys.~Rev.~D {\bf76}, 033006 (2007). 

\bibitem{icepaper} M.~Ackermann {\sl et al.}, J.~Geophys.~Res.~{\bf111}, D13203 (2006).

\bibitem{amandareco} J.~Ahrens {\sl et al.}, Nucl.~Inst.~Meth.~A {\bf524}, 169 (2004).

\bibitem{amanda5yr} A.~ Achterberg \emph{et al.}, Phys.~Rev.~D {\bf75}, 102001 (2007).  

\bibitem{till} T.~ Neunh\"offer, Astropart.~Phys.~{\bf25}, 220 (2006).

\bibitem{bayes} T.~DeYoung \textit{et al.}, in \textit{Advanced Statistical
  Techniques in Particle Physics} (Durham, U.K.), 235 (2002).

\bibitem{amanda7yr} R.~Abbasi \textit{et al.}, Phys.~Rev.~D {\bf79}, 062001 (2009).

\bibitem{nusim} G.~C.~Hill,  Astropart.~Phys.~\textbf{6}, 215 (1997). 

\bibitem{bartol} G.~D.~Barr, T.~K.~Gaisser, P.~Lipari, S.~Robbins, and
T.~Stanev, Phys.~Rev.~D {\bf70}, 023006 (2004). 

\bibitem{gaisserbook} T.~K.~Gaisser, \emph{Cosmic Rays and Particle
  Physics} (Cambridge University, U.K., 1991), p.~88.

\bibitem{mmc} D.~Chirkin and W.~Rhode, arXiv:hep-ph/0407075.

\bibitem{photonics} J.~Lundberg {\em et al.}, Nucl.~Instrum.~Meth.~A
  {\bf581}, 619 (2007).

\bibitem{amasim} S.~Hundertmark, in Proc.~of the Workshop on Simulation and
  Analysis Methods for Large Neutrino Telescopes, DESY-PROC-1999-01, 276
  (1999); S.~Hundertmark, Ph.D.~thesis, Humboldt Universit\"at zu Berlin
  (1999). 

\bibitem{corsika} D.~Heck {\em et al.}, Tech.~Rep.~FZKA 6019,
  Forschungszentrum Karlsruhe (1998). 

\bibitem{fc} G.~J.~Feldman and R.~D.~Cousins, Phys.~Rev.~{\bfseries
  D57}, 3873 (1998).

\bibitem{cousins} R.~Cousins, in \textit{Statistical Problems in Particle
  Physics, Astrophysics and Cosmology: Proceedings of PHYSTAT05}, edited by
  L.~Lyons and M.~\"Unel (Univ.~of Oxford, U.K., 2005).

\bibitem{kendall} A.~Stuart, K.~Ord, and S.~Arnold, \emph{Kendall's
  Advanced Theory of Statistics} (Arnold, London, 1999), vol.~2A, 6th and
  earlier editions. 

\bibitem{minuit} F.~James and M.~Roos, Comp.~Phys.~Comm.~{\bf10}, 343
  (1975). 

\bibitem{rolke2000} W.~A.~Rolke and A.~M.~L\'{o}pez, Nucl.~Instrum.~Meth.~A
  {\bfseries 458}, 745 (2001).

\bibitem{rolke2004} W.~A.~Rolke, A.~M.~L\'{o}pez, and J.~Conrad, in
  \textit{Statistical Problems in Particle Physics, Astrophysics and
    Cosmology: Proceedings of PHYSTAT05}, edited by 
  L.~Lyons and M.~\"Unel (Univ.~of Oxford, U.K., 2005); arXiv:physics/0403059. 

\bibitem{feldman_pc} G.~J.~Feldman, ``Multiple measurements and 
parameters in the unified approach,'' Workshop on 
Confidence Limits, Fermilab (2000).

\bibitem{cranmer05} K.~Cranmer, in \textit{Statistical Problems in Particle
  Physics, Astrophysics and Cosmology: Proceedings of PHYSTAT05}, edited by
  L.~Lyons and M.~\"Unel (Univ.~of Oxford, U.K., 2005); arXiv:physics/0511028.

\bibitem{anis} M.~Kowalski and A.~Gazizov, Comp.~Phys.~Comm.{\bf172} 3, 203
  (2005). 

\bibitem{cteq5} H.~L.~Lai \textit{et al.}, Eur.~Phys.~J.~C {\bf12}, 375
  (2000). 

\bibitem{mrs} A.~D.~Martin, R.~G.~Roberts, and W.~J.~Stirling,
  Phys.~Lett.~B {\bf354}, 155 (1995).

\bibitem{sibyll} R.~S.~Fletcher, T.~K.~Gaisser, P.~Lipari, and T.~Stanev,
  Phys.~Rev.~D {\bf50}, 5710 (1994); R.~Engel, T.~K.~Gaisser, P.~Lipari,
  and T.~Stanev, in Proc.~of 26th ICRC (Salt Lake City) {\bf1}, 415 (1999). 

\bibitem{epos} K.~Werner, F.-M.~Liu, and T.~Pierog, Phys.~Rev.~C {\bf74}, 044902
  (2006). 

\bibitem{qgs2} S.~Ostapchenko, Phys.~Lett.~B {\bf636}, 40 (2006);
  S.~Ostapchenko, Phys.~Rev.~D {\bf74}, 014026 (2006).

\bibitem{naumov} G.~Fiorentini, A.~Naumov, and F.~L.~Villante,
  Phys.~Lett.~B {\bf510}, 173 (2001); E.~V.~Bugaev \textit{et al.}, Nuovo 
  Cimento C {\bf12}, 41 (1989).

\bibitem{forward_folding} S.~Mizobuchi \textit{et al.}, in Proc.~of 29th
  ICRC (Pune) {\bf 5}, 323 (2005); A.~Djannati-Ata\"{\i} \textit{et al.}, A\&A
  {\bf350}, 17 (1999).

\bibitem{icecube_vli} M.~C.~Gonz\'{a}lez-Garc\'{i}a, F.~Halzen, and
  M.~Maltoni, Phys.~Rev.~D {\bf 71}, 093010 (2005).

\bibitem{maltoni_atm} M.~C.~Gonz\'{a}lez-Garc\'{i}a, M.~Maltoni, and
  J.~Rojo, JHEP {\bf0610}, 075 (2006).

\bibitem{frejus} K.~Daum, W.~Rhode, P.~Bareyre, and R.~Barloutaud,
  Z.~Phys.~C {\bf66}, 417 (1995).

\bibitem{cygnus_ob2} L.~A.~Anchordoqui \emph{et al.}, Phys.~Rev.~D {\bf72},
  065019 (2005).

\bibitem{hooper_cpt} D.~Hooper, D.~Morgan, and E.~Winstanley, Phys.~Rev.~D
  {\bf72}, 065009 (2005). 

\bibitem{mewes}V.~A.~Kosteleck\'y and M.~Mewes, Phys.\ Rev.\ D {\bf
  69}, 016005 (2004). 

\bibitem{amelino-camelia_grb} G.~Amelino-Camelia, Intl.~Jour.~of
  Mod.~Phys.~D {\bf12}, 1633 (2003).

\bibitem{piran} U.~Jacob and T.~Piran,  Nature Phys.~{\bf3}, 87 (2007).

\bibitem{halzen_grb}  M.~C.~Gonz\'{a}lez-Garc\'{i}a and F.~Halzen, JCAP
  {\bf2}, 008 (2007).


\bibitem{sme} D.~Colladay and V.~A.~Kosteleck\'y, Phys.~Rev.~D {\bf
  58}, 116002 (1998).

\bibitem{mewes2}V.~A.~Kosteleck\'y and M.~Mewes, Phys.\ Rev.\ D {\bf
  70}, 031902(R) (2004).

\bibitem{tandem}T.~Katori, V.~A.~Kosteleck\'y, and R.~Tayloe,
  Phys.~Rev.~D {\bf74}, 105009 (2006).  

\bibitem{mavromatos_rev} N.~E.~Mavromatos, Lect.~Notes Phys.~{\bf669}, 245
  (2005).

\bibitem{lindblad} G.~Lindblad, Commun.~Math.~Phys.~{\bf48}, 119 (1976).

\bibitem{benatti} F.~Benatti and R.~Floreanini, JHEP {\bf02}, 032 (2000).

\end{thebibliography}
\end{document}